\documentclass[twocolumn,english]{revtex4}
\usepackage[T1]{fontenc}
\usepackage[latin9]{inputenc}
\usepackage{color}
\usepackage{array}
\usepackage{float}
\usepackage{textcomp}
\usepackage{multirow}
\usepackage{amsmath}
\usepackage{amssymb}
\usepackage{graphicx}
\usepackage{wasysym}

\makeatletter

\newcommand*\LyXZeroWidthSpace{\hspace{0pt}}
\newcommand{\lyxmathsym}[1]{\ifmmode\begingroup\def\b@ld{bold}
  \text{\ifx\math@version\b@ld\bfseries\fi#1}\endgroup\else#1\fi}

\providecommand{\tabularnewline}{\\}

\@ifundefined{textcolor}{}
{%
 \definecolor{BLACK}{gray}{0}
 \definecolor{WHITE}{gray}{1}
 \definecolor{RED}{rgb}{1,0,0}
 \definecolor{GREEN}{rgb}{0,1,0}
 \definecolor{BLUE}{rgb}{0,0,1}
 \definecolor{CYAN}{cmyk}{1,0,0,0}
 \definecolor{MAGENTA}{cmyk}{0,1,0,0}
 \definecolor{YELLOW}{cmyk}{0,0,1,0}
}

\makeatother

\usepackage{babel}
\begin{document}
\title{DoNOF: an open-source implementation of natural-orbital-functional-based
methods for quantum chemistry}
\author{Mario Piris$^{1,2,3}$ and Ion Mitxelena$^{1}$\bigskip{}
}
\address{$^{1}$Donostia International Physics Center (DIPC), 20018 Donostia,
Euskadi, Spain.\\
$^{2}$Kimika Fakultatea, Euskal Herriko Unibertsitatea (UPV/EHU),
PK 1072, 20080 Donostia, Euskadi, Spain. \\
$^{3}$Basque Foundation for Science (IKERBASQUE), 48013 Bilbao, Euskadi,
Spain.\bigskip{}
}
\email{mario.piris@ehu.eus}

\selectlanguage{english}%
\begin{abstract}
The natural orbital functional theory (NOFT) has emerged as an alternative
formalism to both density functional (DF) and wavefunction methods.
In NOFT, the electronic structure is described in terms of the natural
orbitals (NOs) and their occupation numbers (ONs). The approximate
NOFs have proven to be more accurate than those of the density for
systems with a significant multiconfigurational character, on one
side, and scale better with the number of basis functions than correlated
wavefunction methods, on the other side. A challenging task in NOFT
is to efficiently perform orbital optimization. In this article we
present DoNOF, our open source implementation based on diagonalizations
that allows to obtain the resulting orbitals automatically orthogonal.
The one-particle reduced-density matrix (1RDM) of the ensemble of
pure-spin states provides the proper description of spin multiplets.
The capabilities of the code are tested on the water molecule, namely,
geometry optimization, natural and canonical representations of molecular
orbitals, ionization potential, and electric moments. In DoNOF, the
electron-pair-based NOFs developed in our group PNOF5, PNOF7 and PNOF7s
are implemented. These $\mathcal{JKL}$-only NOFs take into account
most non-dynamic effects plus intrapair-dynamic electron correlation,
but lack a significant part of interpair-dynamic correlation. Correlation
corrections are estimated by the single-reference NOF-MP2 method that
simultaneously calculates static and dynamic electron correlations
taking as a reference the Slater determinant formed with the NOs of
a previous PNOF calculation. The NOF-MP2 method is used to analyze
the potential energy surface (PES) and the binding energy for the
symmetric dissociation of the water molecule, and compare it with
accurate wavefunction-based methods.\bigskip{}

\textit{Preprint submitted to Computer Physics Communications}
\end{abstract}
\maketitle

\section*{PROGRAM SUMMARY}

\noindent {\em Program Title:} DoNOF: Donostia Natural Orbital
Functional Computer Program\\

\noindent {\em Licensing provisions:} GNU GPL version 3\\

\noindent {\em Programming language:} Fortran\\

\noindent {\em Download:} DoNOF can be obtained from the CPC Program
Library, and from its public git repository \url{https://github.com/DoNOF/DoNOFsw}.\\

\noindent {\em Reference Manual:} The DoNOF online manual is available
through the link \url{https://donof.readthedocs.io/}, from where
you can also download it in pdf format.\\

\noindent {\em Nature of problem:} Accurate solutions require a
balanced treatment of static and dynamic electron correlation. Wavefunction-based
methods can handle correctly both types of correlation, however, are
computational expensive and demand prior knowledge of the system.
Conversely, DF-based methods have a relatively low computational cost.
Nowadays, we can find in the literature a wide range of empirical
or non-empirical parametrized DFs, however, the list of challenges
is still large for DF theory. We need a one-particle formalism more
accurate than approximate DFs, but less computationally demanding
than wavefunction methods. The answer may be to develop a functional
based on the 1RDM. Approximations of the ground-state energy in terms
of NOs and their ONs have been developed for practical applications.
Minimization of the functional is performed under the orthonormality
requirement for the NOs, whereas the ONs conform to the N-representability
conditions for the 1RDM. The main challenge is to efficiently perform
orbital optimization due to the convergence problems that arise when
solving Euler's nonlinear equations. In addition, the $\mathcal{JKL}$-functionals
proposed so far lack a significant part of dynamic correlation, hence
correlation corrections must be added to NOF calculations. \\

\noindent {\em Solution method:} The DoNOF computer program is
designed to solve the energy minimization problem of a NOF that describes
the ground-state of an N-electron system at absolute zero temperature.
The program includes the NOFs developed in the Donostia quantum chemistry
group, namely PNOF5, PNOF7, and PNOF7s. The solution is established
by optimizing the energy functional with respect to the ONs and to
the NOs, separately. The conjugate gradient or limited memory Broyden-Fletcher-Goldfarb-Shanno
(L-BFGS) methods can be employed for the optimization of the ONs.
The optimal NOs are obtained by iterative diagonalizations of a symmetric
matrix {[}1{]} determined by the Lagrange multipliers associated with
the orthogonality conditions of NOs. An aufbau principle makes the
first-order energy contribution negative upon a diagonalization. To
assist convergence, the direct inversion of the iterative subspace
(DIIS) extrapolation technique is used, and a variable scale factor
balances the symmetric matrix. DoNOF also allows subsequent corrections
of the electron correlation once the NOF solution has been obtained.
Corrections can be estimated by the single-reference NOF-MP2 method
{[}2{]} that simultaneously calculates static and dynamic electron
correlations taking as a reference the Slater determinant formed with
the NOs of the previous PNOF solution. The procedure is simple, showing
a formal scaling of $\mathrm{N}_{B}^{5}$ ($\mathrm{N}_{B}$: number
of basis functions).\\

\noindent {\em Restrictions:} DoNOF is intended for non-relativistic
Hamiltonians that does not contain spin coordinates, therefore, our
code cannot handle Hamiltonians which break spin symmetry, for example,
those that contain a magnetic field term.\\

\noindent {\em Unusual features:}\\

\noindent In case of multiplets ($S\neq0$), our implementation differs
from procedures routinely used in electronic structure calculations
that focus on the high-spin component or break the spin symmetry.
For a given spin $S$, we consider a nondegenerate mixed quantum state
that allows all possible $S_{z}$ values.\\

\noindent \textcolor{black}{No prior knowledge of the system is required,
so the method can be employed as a black-box. In addition, PNOF is
parameter free, so the performance is not subject to specific systems
or parametrization data.}\\

\noindent Analytic energy gradients with respect to nuclear motion
are available in NOFT {[}3{]}, so the computation of NOF gradients
in our code is analogous to gradient calculations at the Hartree-Fock
(HF) level of theory. Our implementation allows the calculation of
gradients by simple evaluation without resorting to the linear-response
theory with the corresponding savings of computational time.\\

\noindent {\footnotesize{}{[}1{]} M. Piris, J. M. Ugalde, J. Comput.
Chem. 30 (2009) 2078-2086.}{\footnotesize\par}

\noindent {\footnotesize{}{[}2{]} M. Piris, Phys. Rev. Lett. 119 (2017)
063002; Phys. Rev. A 98 (2018) 022504.}{\footnotesize\par}

\noindent {\footnotesize{}{[}3{]} I. Mitxelena, M. Piris, J. Chem.
Phys. 146 (2017) 014102.}{\footnotesize\par}

\section{\label{sec:Introduction}Introduction}

Today, computational chemistry helps the experimental chemist understand
experimental data, explore reaction mechanisms or predict completely
new molecules. There is no doubt that understanding at the molecular
level will ultimately lead to an \textit{ab initio} process design.
The best solutions to the many-body problem are offered by methods
based on approximate wavefunctions, however, such techniques demand
significant computational resources as soon as the systems grow.

It has been known for a long time \citep{Husimi1940} that the energy
of a system with, at most, two-body interactions is an exact functional
of the two-particle reduced density matrix (2RDM). Realistic variational
2RDM calculations are possible nowadays \citep{Mazziotti2012} but
they are still computationally expensive. The need for treatments
that scale favorably with the number of electrons is evident. The
one-particle functional theories, where the ground-state energy is
represented in terms of the first-order reduced density matrix (1RDM)
\citep{Gilbert1975,Valone1980} or simply the density \citep{Hohenberg1964},
satisfy this requirement. Unfortunately, computational schemes \citep{Valone1991,Mori-Sanchez2018}
based on exact functionals \citep{Levy1979,Lieb1983} are also expensive,
therefore, approximations of the ground-state energy in terms of the
density \citep{Burke2012,Becke2014} or 1RDM \citep{Pernal2016} have
been developed for practical applications.

Usually, the local density and generalized gradient approximations
are employed for density functionals (DFs), however, they are unable
to describe the effects of strong electron correlations. A practical
way around this problem has been to use hybrid-exchange functionals,
or performing a range separation of the interaction. The resulting
functionals appear to work better but the list of failures is still
large, for instance, over stabilization of radicals, wrong description
of activation barriers, poor treatment of charge transfer processes,
and the inability to account for weak interactions \citep{Cohen2012}.
Today we have a large number of empirical and non-empirical parameterized
functionals, but this large-scale parameterization has led to a very
specific applicability in systems and phenomena. For each new challenge,
DFs have to be calibrated against the standard wavefunction methods.
The last strategy of double hybrid and RPA-based functionals represents
a step forward in terms of accuracy, but also leads to a high computational
cost.

The key to overcoming these drawbacks may be to ensure the N-repre-
sentability \citep{Coleman1963}, which is twofold in one-particle
theories \citep{Piris2017a}. First, we have the N-representability
of the fundamental variable, i.e., the density or 1RDM must come from
an N-particle density matrix. Fortunately, conditions to ensure this
have been well-established and are easily implementable. Second, approximate
functionals must reconstruct the known functional E{[}2RDM{]}, so
the functional N-representability problem arises, that is, we have
to meet the requirement that the 2RDM reconstructed in terms of 1RDM
or density must satisfy its N-representability conditions as well.
Most of the approximate functionals currently-in-use are not N-representable
\citep{Ayers2007,Ludena2013,RodriguezMayorga2017}. The N-representability
constraints for acceptable 1RDM or density are easy to implement,
but are insufficient to guarantee that the reconstructed 2RDM is N-representable,
and thereby the approximate functional either.

The unknown functional in a 1RDM-based theory only needs to reconstruct
the electron-electron potential energy ($V_{ee}$). This reflects
an undeniable advantage of 1RDM approximations with respect to approximate
DFs which also require a reconstruction of the one-particle term,
including the kinetic energy. Our goal is to accomplish a 1RDM formalism
more accurate than the approximate DFs, but less computationally demanding
than methods based upon approximate wavefunctions or the 2RDM. In
2005 \citep{Piris2006}, the 2RDM was reconstructed in terms of two-index
auxiliary matrices. The latter satisfy the general symmetry properties
and sum rules, but also known necessary N-representability conditions
of the 2RDM \citep{Piris2010a}, namely, the so-called D, Q and G
positivity conditions.

In the absence of an external field, the non-relativistic Hamiltonian
used in electronic calculations does not contain spin coordinates.
Consequently, the ground state of a many-electron system with total
spin $S$ is a multiplet, i.e., a mixed quantum state that allows
all possible $S_{z}$ values. In this vein, another issue with DFs
is that even within the spin-dependent-density formalism they cannot
describe the degeneracy of the spin-multiplet components. The proper
description of the ensemble of pure spin states is provided by the
1RDM functional theory. In our implementation, the total electronic
spin is imposed, not just the spin projection \citep{Piris2009,Quintero-Monsebaiz2019,Piris2019}.

The reconstruction of the 2RDM can be conveniently done in the natural
orbital representation where the 1RDM is diagonal. In this spectral
representation of the 1RDM, the energy is clearly called natural orbital
functional (NOF). A complete account of the formulation and development
of NOF theory can be found elsewhere \citep{Piris2007}. Different
ways of approximating the aforementioned auxiliary matrices have led
to the appearance of different versions of the Piris NOF (PNOF) \citep{Piris2006,Piris2013b,Piris2014a}.
Remarkably, the performance of these functionals has achieved chemical
accuracy in many cases \citep{Lopez2010}. These functionals are able
to yield \citep{Ruiperez2013} a correct description of systems with
a multiconfigurational nature, one of the biggest challenges for DFs. 

So far, only the NOFs that satisfy electron pairing restrictions \citep{Piris2018a}
are able to provide the correct number of electrons in the fragments
after a homolytic dissociation \citep{Matxain2011}. For instance,
the dissociation curve of the carbon dimer closely resembles that
obtained from the optimized CASSCF(8,8) wavefunction \citep{Piris2016a}.
It has recently been shown \citep{Mitxelena2020,Mitxelena2020a} that
PNOF7 \citep{Piris2017,mitxelena2018a} is an efficient method for
strongly correlated electrons in one and two dimensions. Our implementation
is precisely designed to solve the energy minimization problem of
an electron-pairing-based NOF, presented briefly in section \ref{sec:NOF}.

The solution is established optimizing the energy functional with
respect to the occupation numbers (ONs) and to the natural orbitals
(NOs), separately. Under pairing restrictions, the constrained nonlinear
programming problem for the ONs can be treated as an unconstrained
optimization, with the corresponding saving of computational times.
The orbital optimization is the bottleneck of this algorithm since
direct minimization of the orbitals has been proved to be a costly
method \citep{Cohen2002,Herbert2003a}. In 2009 \citep{Piris2009a},
a self-consistent procedure was proposed which yields the NOs automatically
orthogonal. This scheme requires computational times that scale as
in the Hartree-Fock (HF) approximation. However, our implementation
in the molecular basis set requires also four-index transformation
of the electron repulsion integrals (ERIs), which is the time-consuming
step, though a parallel implementation of this part of the code substantially
improves the performance of our program \citep{Matito2013}. To achieve
convergence, the direct inversion of the iterative subspace (DIIS)
extrapolation technique \citep{Pulay1980,Pulay1982} is used, and
a variable scale factor balances the symmetric matrix subject to the
iterative diagonalizations. In section \ref{sec: Energy-minimization},
our self-consistent algorithm is introduced, and remarks specific
to convergence techniques are discussed. An overview of the DoNOF
computational procedure is given at the end of this section.

Section \ref{sec: MolPro} is dedicated to molecular properties that
can be calculated with DoNOF once the solution is reached. Among these
properties are vertical ionization potentials \citep{Piris2012},
Mulliken population analysis \citep{Mulliken1955}, electric dipole,
quadrupole and octupole moments \citep{Mitxelena2016}, canonical
representation of molecular orbitals \citep{Piris2013}, analytical
energy gradients with respect to nuclear motion \citep{Mitxelena2017},
and geometry optimization \citep{MITXELENA2019}. The program also
provides RDMs at the atomic and molecular bases, and a wave function
file (WFN) to perform additional quantitative and visual analyzes
of the molecular systems.

The fact that electron-pairing-based NOFs proposed so far partially
lack correlation constitutes their major limitation. They do not fully
recover dynamical correlation, which is crucial for the correct description
of potential energy surfaces and dispersion interactions. Recently
\citep{Piris2017,Piris2018b,Piris2019}, a new technique that combines
the NOF theory with the second-order Møller-Plesset (MP2) perturbation
theory was proposed, giving rise to the NOF-MP2 method. The latter
is a single-reference method capable of achieving a balanced treatment
of static and dynamic electron correlation even for those systems
with significant multiconfigurational character. The orbital-invariant
formulation of the NOF-MP2 method is briefly presented in section
\ref{sec: DynCorr}. The method scales formally as $\mathrm{N}_{B}^{5}$,
where $\mathrm{N}_{B}$ is the number of basis functions. The absolute
energies improve over PNOF values and get closer to the values obtained
by accurate wavefunction-based methods \citep{Piris2018b}. Likewise,
NOF-MP2 is able to give \citep{Lopez2019} a quantitative agreement
for dissociation energies, with a performance comparable to that of
the accurate CASPT2 method.

In section \ref{sec: Examples}, we show the capabilities of DoNOF
for the water molecule as an illustrative example. This includes analysis
of molecular orbitals, ionization potential, electric moments, optimized
geometry, harmonic vibrational frequencies, binding energy, and potential
energy surface of the $H_{2}O$ symmetric dissociation. A summary
is given in section \ref{sec: summary}. Atomic units are used throughout
this work.

\section{\label{sec:NOF}Electron-pairing-based NOF for Multiplets}

Let us consider a non-relativistic N-electron Hamiltonian that does
not contain spin coordinates, namely,
\begin{equation}
\hat{H}=\sum\limits _{ik}\mathcal{H}_{ki}\hat{a}_{k}^{\dagger}\hat{a}_{i}+\frac{1}{2}\sum\limits _{ijkl}\left\langle kl|ij\right\rangle \hat{a}_{k}^{\dagger}\hat{a}_{l}^{\dagger}\hat{a}_{j}\hat{a}_{i}\label{Ham}
\end{equation}

In Eq. (\ref{Ham}), $\mathcal{H}_{ki}$ denote the matrix elements
of the one-particle part of the Hamiltonian involving the kinetic
energy and the potential energy operators, and $\left\langle kl|ij\right\rangle $
are the two-particle interaction matrix elements. $\hat{a}_{i}^{\dagger}$
and $\hat{a}_{i}$ are the familiar fermion creation and annihilation
operators associated with the complete orthonormal spin-orbital set
$\left\{ \left|i\right\rangle \right\} =\left\{ \left|p\sigma\right\rangle \right\} $,
where $\sigma$ is used for $\alpha$ and $\beta$ spins. In this
context, the ground state with total spin $S$ is a multiplet, i.e.,
a mixed quantum state (ensemble) that allows all possible spin projections.
For a given $S$, there are $\left(2S+1\right)$ energy degenerate
eigenvectors $\left|SM_{s}\right\rangle $, so the ground state is
defined by the N-particle density matrix statistical operator of all
equiprobable pure states:
\begin{equation}
\mathfrak{\hat{D}}={\displaystyle \dfrac{1}{2S+1}{\displaystyle {\textstyle {\displaystyle \sum_{M_{s}=-S}^{S}}}}\left|SM_{s}\right\rangle \left\langle SM_{s}\right|}\label{DM}
\end{equation}

From Eq. (\ref{Ham}), it follows that the electronic energy is an
exactly and explicitly known functional of the RDMs, 
\begin{equation}
E=\sum\limits _{ik}\mathcal{H}_{ki}\Gamma_{ki}+\sum\limits _{ijkl}\left\langle kl|ij\right\rangle D_{kl,ij}\label{Energy}
\end{equation}
where the 1RDM and 2RDM are
\begin{equation}
\begin{array}{c}
\Gamma_{ki}={\displaystyle \dfrac{1}{2S+1}{\textstyle {\displaystyle \sum_{M_{s}=-S}^{S}}}}\left\langle SM_{s}\right|\hat{a}_{k}^{\dagger}\hat{a}_{i}\left|SM_{s}\right\rangle \\
D_{kl,ij}={\displaystyle {\textstyle {\displaystyle \dfrac{1}{2\left(2S+1\right)}\sum_{M_{s}=-S}^{S}}}}\left\langle SM_{s}\right|\hat{a}_{k}^{\dagger}\hat{a}_{l}^{\dagger}\hat{a}_{j}\hat{a}_{i}\left|SM_{s}\right\rangle 
\end{array}
\end{equation}
We shall use the Löwdin's normalization, in which the traces of the
1RDM and 2RDM are equal to the number of electrons and the number
of electron pairs, respectively. 

The first term of Eq. (\ref{Energy}) is exactly described as a functional
of $\Gamma$, whereas the second is $V_{ee}\left[D\right]$, an explicit
functional of the 2RDM. To construct the functional $V_{ee}\left[\Gamma\right]$,
we employ the representation where the 1RDM is diagonal ($\Gamma_{ki}=n_{i}\delta_{ki}$).
Restriction on the ONs to the range $0\leq n_{i}\leq1$ represents
a necessary and sufficient condition for ensemble N-representability
of the 1RDM \citep{Coleman1963}. This leads to a NOF, namely,
\begin{equation}
E=\sum\limits _{i}n_{i}\mathcal{H}_{ii}+\sum\limits _{ijkl}D[n_{i},n_{j},n_{k},n_{l}]\left\langle kl|ij\right\rangle \label{ENOF}
\end{equation}

In Eq. (\ref{ENOF}), $D[n_{i},n_{j},n_{k},n_{l}]$ represents the
reconstructed ensemble 2RDM from the ONs. For $\hat{S}_{z}$ eigenvectors,
density matrix blocks that conserve the number of each spin type are
non-vanishing, however, only three of them are independent, namely
$D^{\alpha\alpha}$, $D^{\alpha\beta}$, and $D^{\beta\beta}$ \citep{Piris2007}.
In what follows, we briefly describe how we do the reconstruction
of $D$ to achieve an electron-pairing-based NOF for spin-multiplets.
A more detailed description can be found in Ref. \citep{Piris2019}. 

Let us consider that $\mathrm{N_{I}}$ single electrons determine
the spin $S$ of the system, and the rest of electrons ($\mathrm{N_{II}}=\mathrm{N-N_{I}}$)
are spin-paired, so that all spins corresponding to $\mathrm{N_{II}}$
electrons provide a zero spin. Obviously, in the absence of single
electrons ($\mathrm{N_{I}}=0$), the energy (\ref{ENOF}) must be
reduced to a NOF that describes a singlet state.

We focus on the mixed state of highest multiplicity: $2S+1=\mathrm{N_{I}}+1,\,S=\mathrm{N_{I}}/2$.
Then, the expected value of $\hat{S}_{z}$ for the whole ensemble
$\left\{ \left|SM_{s}\right\rangle \right\} $ is zero, namely, 
\begin{equation}
\mathrm{<}\hat{S}_{z}\mathrm{>}=\frac{1}{\mathrm{N_{I}}+1}{\textstyle {\displaystyle \sum_{M_{s}=-\mathrm{N_{I}}/2}^{\mathrm{N_{I}}/2}}M_{s}}=0\label{Sz0}
\end{equation}

According to Eq. (\ref{Sz0}), we can adopt the spin-restricted theory
in which a single set of orbitals $\left\{ \left|p\right\rangle \right\} $
is used for $\alpha$ and $\beta$ spins. All spatial orbitals will
be then double occupied in the ensemble, so that occupancies for particles
with $\alpha$ and $\beta$ spins are equal: $n_{p}^{\alpha}=n_{p}^{\beta}=n_{p}.$

In turn let us divide the orbital space $\Omega$ into two subspaces:
$\Omega=\Omega_{\mathrm{I}}\oplus\Omega_{\mathrm{II}}$. $\Omega_{\mathrm{II}}$
is composed of $\mathrm{N_{II}}/2$ mutually disjoint subspaces $\Omega{}_{g}$.
Each subspace $\Omega{}_{g}\in\Omega_{\mathrm{II}}$ contains one
orbital $\left|g\right\rangle $ with $g\leq\mathrm{N_{II}}/2$, and
$\mathrm{N}_{g}$ orbitals $\left|p\right\rangle $ with $p>\mathrm{N_{II}}/2$,
namely,
\begin{equation}
\Omega{}_{g}=\left\{ \left|g\right\rangle ,\left|p_{1}\right\rangle ,\left|p_{2}\right\rangle ,...,\left|p_{\mathrm{N}_{g}}\right\rangle \right\} \label{OmegaG}
\end{equation}

Taking into account the spin, the total occupancy for a given subspace
$\Omega{}_{g}$ is 2, which is reflected in the following sum rule:
\begin{equation}
n_{g}+\sum_{i=1}^{\mathrm{N}_{g}}n_{p_{i}}=1,\quad g=1,2,...,\frac{\mathrm{N_{II}}}{2}\label{sum1}
\end{equation}

In general, $\mathrm{N}_{g}$ may be different for each subspace,
but it should be sufficient for the description of each electron pair.
In our implementation, $\mathrm{N}_{g}$ is equal to a fixed number
for all subspaces $\Omega{}_{g}\in\Omega_{\mathrm{II}}$. The maximum
possible value of $\mathrm{N}_{g}$ is determined by the basis set
used in calculations. 

From (\ref{sum1}), it follows that
\begin{equation}
2\sum_{p\in\Omega_{\mathrm{II}}}n_{p}=2\sum_{g=1}^{\frac{\mathrm{N_{II}}}{2}}\left(n_{g}+\sum_{i=1}^{\mathrm{N}_{g}}n_{p_{i}}\right)=\mathrm{N_{II}}\label{sumNp}
\end{equation}

Here, the notation $p\in\Omega_{\mathrm{II}}$ represents all the
indexes of $\left|p\right\rangle $ orbitals belonging to $\Omega_{\mathrm{II}}$.
It is important to recall that orbitals belonging to each subspace
$\Omega_{g}$ vary along the optimization process until the most favorable
orbital interactions are found. Therefore, the orbitals do not remain
fixed in the optimization process, they adapt to the problem.

\textcolor{black}{}
\begin{figure}
\noindent \begin{centering}
\textcolor{black}{\caption{\label{fig1} Splitting of the orbital space $\Omega$ into subspaces.
In this example, $S=3/2$ (quartet) and $\mathrm{N_{I}}=3$, so three
orbitals make up the subspace $\Omega_{\mathrm{I}}$, whereas four
electrons ($\mathrm{N_{II}}=4$) distributed in two subspaces $\left\{ \Omega_{1},\Omega_{2}\right\} $
make up the subspace $\Omega_{\mathrm{II}}$. Note that $\mathrm{N}_{g}=4$.
The arrows depict the values \LyXZeroWidthSpace \LyXZeroWidthSpace of
the ensemble occupation numbers, alpha ($\downarrow$) or beta ($\uparrow$),
in each orbital.\bigskip{}
}
}
\par\end{centering}
\noindent \centering{}\textcolor{black}{\includegraphics[scale=0.5]{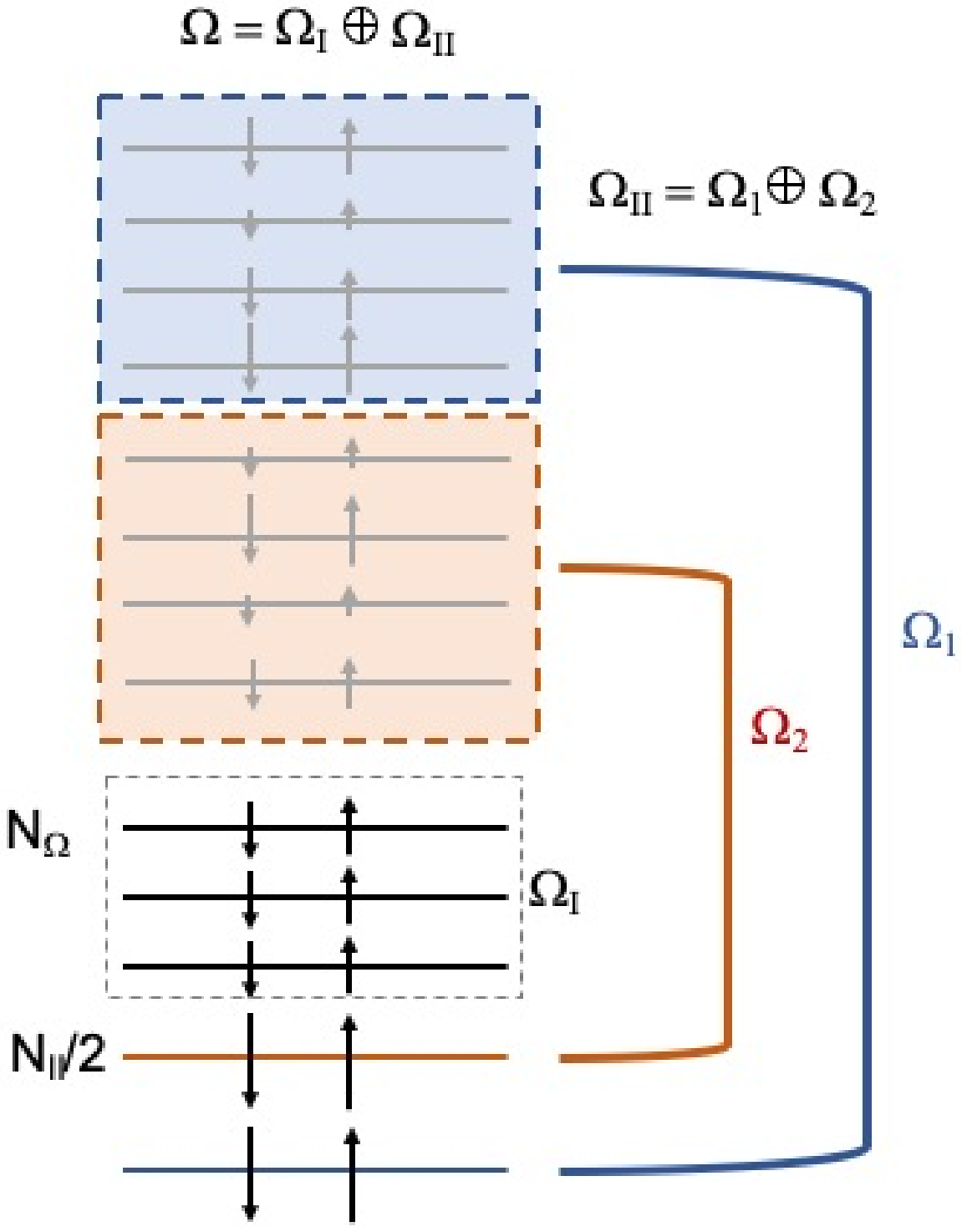}}
\end{figure}

Similarly, $\Omega_{\mathrm{I}}$ is composed of $\mathrm{N_{I}}$
mutually disjoint subspaces $\Omega{}_{g}$. In contrast to $\Omega_{\mathrm{II}}$,
each subspace $\Omega{}_{g}\in\Omega_{\mathrm{I}}$ contains only
one orbital $g$ with $2n_{g}=1$. It is worth noting that each orbital
is completely occupied individually, but we do not know whether the
electron has $\alpha$ or $\beta$ spin: $n_{g}^{\alpha}=n_{g}^{\beta}=n_{g}=1/2$.
It follows that
\begin{equation}
2\sum_{p\in\Omega_{\mathrm{I}}}n_{p}=2\sum_{g=\frac{\mathrm{N_{II}}}{2}+1}^{\mathrm{N}_{\Omega}}n_{g}=\mathrm{N_{I}},
\end{equation}
where $\mathrm{\mathrm{N}_{\Omega}=}\mathrm{N_{II}}/2+\mathrm{N_{I}}$
denotes the total number of subspaces in $\Omega$. Taking into account
Eq. (\ref{sumNp}), the trace of the 1RDM is verified equal to the
number of electrons:
\begin{equation}
2\sum_{p\in\Omega}n_{p}=2\sum_{p\in\Omega_{\mathrm{II}}}n_{p}+2\sum_{p\in\Omega_{\mathrm{I}}}n_{p}=\mathrm{N_{II}}+\mathrm{N_{I}}=\mathrm{\mathrm{N}}
\end{equation}

In Fig. \ref{fig1}, an illustrative example is shown. In this example,
$\mathrm{N_{I}}=3$ ($S=3/2$) hence three orbitals make up the subspace
$\Omega_{\mathrm{I}}$, whereas four electrons ($\mathrm{N_{II}}=4$)
distributed in two subspaces $\left\{ \Omega_{1},\Omega_{2}\right\} $
make up the subspace $\Omega_{\mathrm{II}}$. In Fig. \ref{fig1},
$\mathrm{N}_{g}=4$ corresponds to the maximum allowed value.

In DoNOF, we employ a two-index reconstruction $D[n_{p},n_{q}]$ \citep{Piris2006},
instead of the general dependence on four indices. The necessary N-representability
D, Q, and G conditions of the 2RDM, also known as (2,2)-positivity
conditions \citep{Mazziotti2012a}, impose strict inequalities on
$D$-elements \citep{Piris2010a}. Given these constraints, some proposals
have resulted in several $\mathcal{JKL}$-only functionals \citep{Piris2013b},
where $\mathcal{J}$ and $\mathcal{K}$ refer to the usual Coulomb
and exchange integrals, while $\mathcal{L}$ denotes the exchange-time-inversion
integral \citep{Piris1999}.

For an electron-pairing-based NOF, we divide the matrix elements of
$D$ into intra- and inter-subspace contributions. The intra-subspace
blocks only involves intrapair $\alpha\beta$-contributions of orbitals
belonging to $\Omega_{\mathrm{II}}$. Actually, there can be no interactions
between electrons with opposite spins in a singly occupied orbital,
since there is only one electron with $\alpha$ or $\beta$ spin in
each $\left|SM\right\rangle $ of the ensemble. In the simplest case
of two electrons, an accurate NOF is well-known from the exact wavefunction
\citet{Lowdin1955d}. Consequently, the intrapair $\alpha\beta$-contributions
are

\noindent 
\begin{equation}
\begin{array}{c}
D_{pq,rt}^{\alpha\beta}={\displaystyle \frac{\Pi_{pr}}{2}}\delta_{pq}\delta_{rt}\delta_{p\Omega_{g}}\delta_{r\Omega_{g}}\;(g\leq\frac{N_{\mathrm{II}}}{2})\quad\\
\\
\Pi_{pr}=\left\{ \begin{array}{c}
\sqrt{n_{p}n_{r}}\qquad p=r\textrm{ or }p,r>\frac{N_{\mathrm{II}}}{2}\\
-\sqrt{n_{p}n_{r}}\qquad p=g\textrm{ or }r=g\qquad\;
\end{array}\right.\\
\\
\delta_{p\Omega_{g}}=\left\{ \begin{array}{c}
1\qquad p\in\Omega_{g}\in\Omega_{\mathrm{II}}\\
0\qquad p\notin\Omega_{g}\in\Omega_{\mathrm{II}}
\end{array}\right.
\end{array}\label{intra}
\end{equation}
Note that $D_{pp,pp}^{\alpha\beta}=0$, $\forall p\in\Omega_{\mathrm{I}}$.
For inter-subspace contributions ($\Omega_{f}\neq\Omega{}_{g}$),
the spin-parallel matrix elements are HF like, namely,
\begin{equation}
D_{pq,rt}^{\sigma\sigma}={\displaystyle \frac{n_{p}n_{q}}{2}}\left(\delta_{pr}\delta_{qt}-\delta_{pt}\delta_{qr}\right)\delta_{p\Omega_{f}}\delta_{q\Omega_{g}},\:{}_{\sigma=\alpha,\beta}\label{interaa}
\end{equation}
whereas the spin-anti-parallel blocks are
\begin{equation}
\begin{array}{c}
D_{pq,rt}^{\alpha\beta}={\displaystyle {\displaystyle \frac{n_{p}n_{q}}{2}}}\delta_{pr}\delta_{qt}\delta_{p\Omega_{f}}\delta_{q\Omega_{g}}-{\displaystyle \frac{\Phi_{p}\Phi_{r}}{2}}\delta_{p\Omega_{f}}\delta_{r\Omega_{g}}\cdot\\
\\
\cdot\left\{ \begin{array}{c}
\delta_{pq}\delta_{rt}\quad f\leq\frac{\mathrm{N_{II}}}{2}\textrm{ or }g\leq\frac{\mathrm{N_{II}}}{2}\\
\quad\delta_{pt}\delta_{qr}\qquad\frac{\mathrm{N_{II}}}{2}<f,g\leq\mathrm{N}_{\Omega}\quad
\end{array}\right.
\end{array}\label{interab}
\end{equation}

\noindent where $\Phi_{p}=\sqrt{n_{p}(1-n_{p})}$. It is not difficult
to verify \citep{Piris2019} that the reconstruction (\ref{intra})-(\ref{interab})
leads to $\mathrm{<}\hat{S}^{2}\mathrm{>}=S\left(S+1\right)$ with
$S=\mathrm{N_{I}}/2$.

In quantum chemistry, we usually use real spatial orbitals, then $\mathcal{L}_{pq}=\mathcal{K}_{pq}$.
After a simple algebra, the energy (\ref{ENOF}) can be written as
\begin{equation}
E=\sum\limits _{g=1}^{\frac{\mathrm{N_{II}}}{2}}E_{g}+\sum_{g=\frac{\mathrm{N_{II}}}{2}+1}^{\mathrm{N}_{\Omega}}\mathcal{H}_{gg}+\sum\limits _{f,g=1;f\neq g}^{\mathrm{N}_{\Omega}}E_{fg}\label{EPNOF7}
\end{equation}
where
\begin{equation}
E_{g}=2\sum\limits _{p\in\Omega_{g}}n_{p}\mathcal{H}_{pp}+\sum\limits _{q,p\in\Omega_{g}}\Pi_{qp}\mathcal{K}_{pq}\,,\;\Omega{}_{g}\in\Omega_{\mathrm{II}}\label{Eg}
\end{equation}
is the energy of an electron pair with opposite spins. $E_{fg}$ correlates
the motion of electrons with parallel and opposite spins belonging
to different subspaces ($\Omega_{f}\neq\Omega{}_{g}$):
\begin{equation}
E_{fg}=\sum\limits _{p\in\Omega_{f}}\sum\limits _{q\in\Omega_{g}}\left[n_{q}n_{p}\left(2\mathcal{J}_{pq}-\mathcal{K}_{pq}\right)-\Phi_{q}\Phi_{p}\mathcal{K}_{pq}\right]\label{Efg}
\end{equation}
The functional (\ref{EPNOF7})-(\ref{Efg}) is PNOF7 for multiplets
\citep{Piris2017,mitxelena2018a,Piris2018b}. It is
worth noting that PNOF7 is reduced to PNOF5 \citep{Piris2011,Piris2013e}
for $\Phi_{p}\equiv0$. The latter constitutes an independent-pair
approximation by not considering the interaction between orbitals
that belong to different subspaces beyond the HF terms of Eq. (\ref{Efg}).
It will afford good results when the number of electron pairs is small,
but the results deteriorate rapidly as the system grows. Interestingly,
an antisymmetrized product of strongly orthogonal geminals (APSG)
with the expansion coefficients explicitly expressed by the ONs also
leads to PNOF5 \citep{Piris2013e} showing that PNOF5 is strictly
N-representable. As far as we know, there is no other NOF for which
its generating wavefunction is known.

PNOF7 introduces interaction terms between orbitals
belonging to different subspaces. Its main weakness is the absence
of the dynamic electron correlation between pairs, since $\Phi_{p}=\sqrt{n_{p}(1-n_{p})}$
has significant values only when the ONs differ substantially from
1 and 0. Consequently, PNOF7 is able to recover the complete intrapair,
but only the static interpair correlation. The functional has proven
its worth in problems where a strong correlation is revealed. We have
recently demonstrated \citep{Mitxelena2020,Mitxelena2020a} the ability
of PNOF7 to describe strong correlation effects in one-dimensional
and two-dimensional systems by comparing our results with exact diagonalization,
density matrix renormalization group, and quantum Monte Carlo calculations.
Unfortunately, PNOF7 does not include dynamic interpair correlation,
which can be included a posteriori using a modified version of standard
second-order perturbation theory (see below in section \ref{sec: DynCorr}).

\section{\label{sec: Energy-minimization} Energy minimization problem}

Minimization of the energy functional $E\left[\mathrm{N},\left\{ n_{p}\right\} ,\left\{ \varphi_{p}\left(\mathbf{r}\right)\right\} \right]$
is performed under the orthonormality requirement for the real spatial
orbitals,
\begin{equation}
\left\langle p|q\right\rangle =\int d\mathbf{r}\varphi_{p}\left(\mathbf{r}\right)\varphi_{q}\left(\mathbf{r}\right)=\delta_{pq}\label{ortho}
\end{equation}

\noindent whereas the occupancies conform to the ensemble N-representability
conditions $0\leq n_{p}\leq1$, and pairing sum rules (\ref{sum1}).
The latter are important for orbitals belonging to $\Omega_{\mathrm{II}}$
since $n_{p}=1/2$, $\forall p\in\Omega_{\mathrm{I}}$. \textcolor{black}{This
is a constrained optimization problem. The solution is established
by optimizing the energy (}\ref{EPNOF7}\textcolor{black}{) with respect
to the ONs and to the NOs, separately.}

\subsection{Occupancy Optimization}

Bounds on $\left\{ n_{p}\right\} $ can be imposed automatically by
expressing the ONs through new auxiliary variables $\left\{ \gamma_{p}\right\} $,
and using trigonometric functions. Likewise, the equality constrains
(\ref{sum1}) can also always be satisfied using the properties of
these functions. In this way, we transform the constrained minimization
problem of the objective function with respect to $\left\{ n_{p}\right\} $
into the problem of minimizing $E$ with respect to auxiliary variables
$\left\{ \gamma_{p}\right\} $ without restrictions on their values.
In our implementation, \textcolor{black}{the conjugate gradient (CG)
method \citep{Fletcher2000} is used for performing the optimization
of the energy with respect to $\gamma$-variables.}

Let us set the ON of an orbital $\left|g\right\rangle $ as
\begin{equation}
n_{g}=\dfrac{1}{2}\left(1+cos^{2}\gamma_{g}\right);\:g=1,2,...,\mathrm{N_{II}}/2\label{soo}
\end{equation}
Given the range $[0,1]$ of possible values for $cos^{2}\gamma_{g}$,
it is easy to see that $1/2\leq n_{g}\leq1$. Accordingly, we shall
name $\left\{ \varphi_{g}\left(\mathbf{r}\right);g=1,2,...,N_{\mathrm{II}}/2\right\} $
strongly occupied orbitals. The pairing conditions (\ref{sum1}) can
be conveniently written as
\begin{equation}
\sum_{i=1}^{\mathrm{N}_{g}}n_{p_{i}}=h_{g};\quad g=1,2,...,\mathrm{N_{II}}/2\label{sum1-h}
\end{equation}
where the hole $h_{g}=1-n_{g}$ in the orbital $\left|g\right\rangle $
is $h_{g}=\left(sin^{2}\gamma_{g}\right)/2$. Let us express the ONs
of the rest of orbitals through new auxiliary variables while satisfying
Eq. (\ref{sum1-h}). Indeed, the ONs $\left\{ n_{p_{1}},n_{p_{2}},...,n_{p_{\mathrm{N}_{g}}}\right\} $
of each subspace $\Omega{}_{g}\in\Omega_{\mathrm{II}}$ can be set
as
\begin{equation}
\begin{array}{c}
n_{p_{1}}=h_{g}sin^{2}\gamma_{p_{1}}\\
\\
n_{p_{2}}=h_{g}cos^{2}\gamma_{p_{1}}sin^{2}\gamma_{p_{2}}\\
\\
\cdots\\
\\
n_{p_{i}}=h_{g}cos^{2}\gamma_{p_{1}}cos^{2}\gamma_{p_{2}}\cdots cos^{2}\gamma_{p_{i-1}}sin^{2}\gamma_{p_{i}}\\
\\
\cdots\\
\\
n_{p_{\mathrm{N}_{g}-1}}=h_{g}cos^{2}\gamma_{p_{1}}cos^{2}\gamma_{p_{2}}\cdots cos^{2}\gamma_{p_{\mathrm{N}_{g}-2}}sin^{2}\gamma_{p_{\mathrm{N}_{g}-1}}\\
\\
n_{p_{\mathrm{N}_{g}}}=h_{g}cos^{2}\gamma_{p_{1}}cos^{2}\gamma_{p_{2}}\cdots cos^{2}\gamma_{p_{\mathrm{N}_{g}-2}}cos{}^{2}\gamma_{p_{\mathrm{N}_{g}-1}}\\
\\
\end{array}\label{woo}
\end{equation}

It is not difficult to verify if we add the equations of (\ref{woo})
in pairs, starting with the last two and continuing upwards using
the resulting equation of each sum, that the restriction (\ref{sum1-h})
is always fulfilled thanks to the fundamental trigonometric identity.
Note that by eliminating the equality restrictions (\ref{sum1-h})
from the problem, we move from having $\mathrm{N}_{g}$ unknown occupancies
to $\mathrm{N}_{g}-1$ auxiliary $\gamma$-variables. Finally, taking
into account that $sin^{2}\left(\gamma\right)$ and $cos^{2}\left(\gamma\right)$
are always in the interval $[0,1]$, and that these multiply the magnitude
of the hole $h_{g}$, the rest of orbitals in $\Omega_{\mathrm{II}}$
are weakly occupied, i.e., $0\leq n_{p}\leq1/2$ if $p>\mathrm{N_{II}}/2\cap p\in\Omega_{\mathrm{II}}$.

\subsection{Orbital Optimization}

The orthonormal conditions may be taken into account by the Lagrange
multiplier method. For a fixed set of occupancies, let us introduce
the symmetric multipliers $\left\{ \lambda_{pq}\right\} $ associated
with orthonormality constraints (\ref{ortho}) on the real spatial
orbitals $\left|p\right\rangle $, and define the auxiliary functional
$\Omega$ as
\begin{equation}
\Omega=E-2{\displaystyle \sum_{pq}}\lambda_{qp}\left(\left\langle p|q\right\rangle -\delta_{pq}\right)\label{omega}
\end{equation}

The functional (\ref{omega}) has to be stationary with respect to
variations in $\varphi_{p}\left(\mathbf{r}\right)$, that is

\begin{equation}
\delta\Omega=\sum\limits _{p}\int d\mathbf{r}\delta\varphi_{p}\left(\mathbf{r}\right)\left[\frac{\delta E}{\delta\varphi_{p}\left(\mathbf{r}\right)}-{\displaystyle 4\sum_{q}}\lambda_{qp}\varphi_{q}\left(\mathbf{r}\right)\right]=0\label{variation}
\end{equation}

\noindent which leads to the following orbital Euler equations
\begin{equation}
\frac{\delta E}{\delta\varphi_{p}\left(\mathbf{r}\right)}=4n_{p}\mathcal{\hat{H}}\varphi_{p}\left(\mathbf{r}\right)+\frac{\delta V_{ee}}{\delta\varphi_{p}\left(\mathbf{r}\right)}={\displaystyle 4\sum_{q}}\lambda_{qp}\varphi_{q}\left(\mathbf{r}\right)\label{lowdin}
\end{equation}

The functional derivative of $V_{ee}$ with respect to $\varphi_{p}$
at the point $\mathbf{r}$ depends on the subspace $\Omega{}_{g}$
to which the orbital belongs. For orbitals belonging to a subspace
with a single electron ($\varphi_{p}\in\Omega{}_{g}\in\Omega_{\mathrm{I}}$)
only inter-pair contributions appear, and the functional derivative
is given by the following expression

\begin{equation}
\begin{array}{c}
{\displaystyle \frac{\delta V_{ee}^{inter}}{\delta\varphi_{p}\left(\mathbf{r}\right)}}=4\left\{ \sum\limits _{f=1;f\neq g}^{\mathrm{N}_{\Omega}}\sum\limits _{q\in\Omega_{f}}\left[n_{q}n_{p}\left(2\mathcal{\hat{J}}_{q}-\hat{\mathcal{K}}_{q}\right)\right.\right.\\
\\
\left.\left.-\Phi_{q}\Phi_{p}\hat{\mathcal{K}}_{q}\right]\right\} \varphi_{p}\left(\mathbf{r}\right)
\end{array}\label{Vee-s}
\end{equation}
where
\begin{equation}
\mathcal{\hat{J}}_{q}\left(\mathbf{r}\right)=\int d\mathbf{r}'\frac{\;\left|\varphi_{q}\left(\mathbf{r}'\right)\right|^{2}}{\left|\mathbf{\mathbf{r}-r}'\right|}\label{Coul}
\end{equation}

\begin{equation}
\hat{\mathcal{K}}_{q}\left(\mathbf{r}\right)=\int d\mathbf{r}'\varphi_{q}\left(\mathbf{r}'\right)\frac{\hat{P}_{\mathbf{r},\mathbf{r}'}}{\left|\mathbf{\mathbf{r}-r}'\right|}\varphi_{q}\left(\mathbf{r}'\right)\label{Exc}
\end{equation}
are the usual Coulomb and exchange operators, respectively, and $\hat{P}_{\mathbf{r},\mathbf{r}'}$
is the permutation operator. For orbitals belonging to a subspace
with an electron pair we must also include the intrapair contribution,

\begin{equation}
\frac{\delta V_{ee}}{\delta\varphi_{p}\left(\mathbf{r}\right)}=\frac{\delta V_{ee}^{inter}}{\delta\varphi_{p}\left(\mathbf{r}\right)}+4\sum\limits _{q\in\Omega_{g}}\Pi_{qp}\mathcal{\hat{K}}_{q}\varphi_{p}\left(\mathbf{r}\right);\:\varphi_{p}\in\Omega{}_{g}\in\Omega_{\mathrm{II}}\label{Vee-d}
\end{equation}

The functional (\ref{omega}) must also be stationary with respect
to variations in Lagrange multipliers, which leads to the Eqs. (\ref{ortho}).
For a fixed set of ONs, we have to find $\left\{ \varphi_{p}\right\} $
and $\left\{ \lambda_{qp}\right\} $ that solve Eqs. (\ref{ortho})
and (\ref{lowdin}). In general, the energy functional of Eq. (\ref{EPNOF7})
is not invariant with respect to an orthogonal transformation of the
orbitals. Consequently, Eq. (\ref{lowdin}) cannot be reduced to a
pseudo-eigenvalue problem by diagonalizing the $\mathbf{\lambda}$
matrix. This system of equations is nonlinear in $\varphi_{p}$ so
it is formidable to solve, although the symmetry property of $\mathbf{\lambda}$
may be used to simplify the problem. In our implementation, we employ
\textcolor{black}{the self-consistent procedure proposed in \citep{Piris2009a}
to obtain the NOs.}

Multiplying the Eq. (\ref{lowdin}) by $\varphi_{p}$, integrating
over $\mathbf{r}$, and taking into account the Eq. (\ref{ortho}),
we get
\begin{equation}
\lambda_{qp}=n_{p}\mathcal{H}_{qp}+g_{pq}\label{lag}
\end{equation}
where
\begin{equation}
g_{pq}=\int d\mathbf{r}\frac{\delta V_{ee}}{\delta\varphi_{p}\left(\mathbf{r}\right)}\varphi_{q}\left(\mathbf{r}\right)\label{gval}
\end{equation}

At the extremum, the matrix of the Lagrange multipliers must be a
symmetric matrix : $\lambda_{qp}=\lambda_{pq}$. Considering $\lambda_{qp}-\lambda_{pq}$,
and that $\mathcal{H}$ is a symmetric matrix, it follows without
difficulty that
\begin{equation}
\left(n_{p}-n_{q}\right)\mathcal{H}_{qp}+g_{pq}-g_{qp}=0\label{lamlam}
\end{equation}

Eq. (\ref{lamlam}) eliminates the Lagrange multipliers as variables
in the problem, that is, the original Eqs. have been replaced by the
system of Eqs. (\ref{lamlam}) and (\ref{ortho}) over the unknown
$\varphi_{p}$ functions only. Introducing a set of known basis functions,
the integral differential Eqs. (\ref{lamlam}) can be cast into an
algebraic equations, however, the standard methods for solving this
nonlinear system of Eqs. converge very slowly.

Let us define the off-diagonal elements of a symmetric matrix $\mathcal{F}$
as
\begin{equation}
\mathcal{F}_{qp}=\theta\left(p-q\right)[\lambda_{qp}-\lambda_{pq}]+\theta\left(q-p\right)[\lambda_{pq}-\lambda_{qp}]\label{Fockian}
\end{equation}

\noindent where $\theta\left(x\right)$ is the unit-step Heaviside
function. According to Eq. (\ref{lamlam}), $\mathcal{F}_{qp}$ vanishes
at the extremum, hence matrices $\mathcal{F}$ and $\Gamma$ can be
brought simultaneously to a diagonal form at the solution. Hence,
$\varphi_{p}$'s which solve the Eqs. (\ref{lamlam}), may be accomplished
by diagonalization of the matrix $\mathcal{F}$ in an iterative way.
A remarkable advantage of this procedure is that the orthonormality
constraints (\ref{ortho}) are automatically guaranteed. Unfortunately,
the diagonal elements cannot be determined from the symmetry property
of $\mathbf{\lambda}$, so this procedure does not provide a generalized
Fockian in the conventional sense. Nevertheless, $\lambda_{pp}$ may
be determined with the help of an aufbau principle.

Let us consider an \textcolor{black}{almost diagonal} matrix $\mathcal{F}^{0}$
constructed from the set of orbitals $\left\{ \varphi_{p}^{0}\right\} $
according to Eq. (\ref{Fockian}) with diagonal elements $\mathcal{F}_{pp}^{0}$,
and let $E^{0}$ be the corresponding electronic energy. Up to the
first-order terms, the diagonalization of $\mathcal{F}^{0}$ gives
rise to a new set of orbitals \citep{Saunders1973},
\begin{equation}
\varphi_{p}=\varphi_{p}^{0}+\sum_{q,q\neq p}\frac{\mathcal{F}_{qp}^{0}\varphi_{q}^{0}}{\mathcal{F}_{pp}^{0}-\mathcal{F}_{qq}^{0}}\label{fi1}
\end{equation}

The first-order expression for the energy corresponding to the orbitals
defined in Eq. (\ref{fi1}) reads
\begin{equation}
E=E^{0}+2\sum_{p<q}\frac{\left|\mathcal{F}_{pq}^{0}\right|^{2}}{\mathcal{F}_{pp}^{0}-\mathcal{F}_{qq}^{0}}\label{ene1}
\end{equation}

It is clear from Eq. (\ref{ene1}) that if we choose $\mathcal{F}_{qq}^{0}>\mathcal{F}_{pp}^{0}$,
making the first-order energy contribution negative, the energy is
bound to drop upon the diagonalization of $\mathcal{F}^{0}$. Consequently,
NOs may be attained by iterative diagonalization of matrix $\mathcal{F}$
employing the above aufbau principle for the definition of the diagonal
elements. 

In DoNOF, we maintain the values \LyXZeroWidthSpace \LyXZeroWidthSpace of
the previous diagonalization of $\mathcal{F}$ to use them as diagonal
elements in each step. Different $\mathcal{F}$ matrices may come
out depending on the initial guess. In the absence of static correlation,
a good start for $\left\{ \mathcal{F}_{pp}^{0}\right\} $ is usually
the values \LyXZeroWidthSpace \LyXZeroWidthSpace obtained after an
energy optimization with respect to \textcolor{black}{$\gamma$ and
}a single diagonalization of the symmetrized $\lambda$-matrix, $\left(\lambda_{pq}+\lambda_{qp}\right)/2$,
calculated with the HF orbitals\textcolor{black}{. If static correlation
plays an important role, it is generally a better option to replace
the HF orbitals with the eigenvectors of }the one-particle part of
the Hamiltonian ($\mathcal{H}$)\textcolor{black}{.}

From Eq. (\ref{ene1}), it can be seen that the off-diagonal elements
must be small enough, so that each element $\mathcal{F}_{pq}$ communicates
its appropriate significance in each step of the iterative diagonalization
process. Accordingly, a well-scaled $\mathcal{F}$ matrix turns crucial
in order to decrease the energy. In our implementation, the scaling
of $\mathcal{F}$ is achieved by dividing those elements $\mathcal{F}_{pq}$
that exceed some small value $\zeta$ several times by ten, so that
the value of each matrix element results of the same order of magnitude
and lesser than the upper selected bound $\zeta$.

It is important to note that the orbitals belonging
to different subspaces vary throughout the optimization process until
the most favorable orbital interactions are found, that is, there
is no impediment to mixing orbitals from the different subspaces to
arrive at the optimal orbitals. Consequently, the orbital optimization
procedure is independent of the selected initial orbital coupling,
although a proper initial guess favors a faster convergence of this procedure.

\subsection{Convergence Acceleration}

It is well-known that iterative methods frequently suffer from slow
convergence. To accelerate convergence we have implemented a DIIS
extrapolation technique \citep{Pulay1980}. In the latter, an error
vector $\mathbf{e}^{j}$ is constructed at each $j$th step. The construction
of a suitable error vector is related to the gradient of the electronic
energy with respect to $\varphi$ and thus vanishes for the solution.
In DoNOF, we use the off-diagonal elements of the symmetric matrix
$\mathcal{F}$ given by Eq. (\ref{Fockian}) to form the error vector,
namely, $e_{pq}^{j}=\mathcal{F}_{pq}^{j}$ (recall that $\mathcal{F}_{pq}\rightarrow0$
at the converged solution).

Toward the end of the iterative procedure, changes in $\varphi$ are
small and it is possible to find a linear combination of $m$ consecutive
error vectors that approximates the zero vector in the least-squares
sense:
\begin{equation}
\Delta\mathbf{e}=\sum_{j=1}^{m}c_{j}\mathbf{e}^{j}\rightarrow0
\end{equation}
where the coupling constants $c_{j}$ satisfy the normalization condition.
\begin{equation}
\sum_{j=1}^{m}c_{j}=1\label{norma}
\end{equation}

The determination of the coupling constants is achieved by minimizing
the norm of $\Delta\mathbf{e}$
\begin{equation}
\left\Vert \Delta\mathbf{e}\right\Vert =\sum_{i,j=1}^{m}c_{i}c_{j}\left\langle \mathbf{e}^{i}|\mathbf{e}^{j}\right\rangle =\sum_{i,j=1}^{m}c_{i}c_{j}B_{ij}
\end{equation}
under condition (\ref{norma}). This least-squares criterion leads
to the following small set of linear equations:
\begin{equation}
\left(\begin{array}{cc}
\mathbf{B} & -\mathbf{1}\\
-\mathbf{1} & 0
\end{array}\right)\left(\begin{array}{c}
\mathbf{c}\\
\Lambda
\end{array}\right)=\left(\begin{array}{c}
\mathbf{\;0}\\
-1
\end{array}\right)
\end{equation}
where $B_{ij}=\left\langle \mathbf{e}^{i}|\mathbf{e}^{j}\right\rangle $
is a suitably defined metric in the $\mathbf{e}$ space, and $\Lambda$
is a Lagrangian multiplier.

Consequently, the symmetric matrix $\mathcal{F}$ can be estimated
as a linear combination of $m$ previous matrices,

\begin{equation}
\mathcal{\bar{F}}=\sum_{j=1}^{m}c_{j}\mathcal{F}^{j}
\end{equation}

An interpolation like this violates generally the symmetry of $\mathcal{F}$
in second order; however, this becomes insignificant as convergence
is approached. In practice, it has been found that even interpolation
between fairly different matrices gives satisfactory results \citep{Pulay1982}.

Now that all the pieces of the algorithm have been introduced, the
energy optimization procedure implemented in DoNOF can be described.

\subsection{\label{subsec:Procedure}Computational Procedure}

In 2009, the DoNOF code began to be developed from reading the matrix
elements necessary to perform calculations generated by the GAMESS
program \citep{Schmidt1993,Gordon2005}. For these historical reasons,
we keep the GAMESS format to enter the basic molecular information,
including the molecular basis set that can be downloaded from the
Basis Set Exchange \citep{Pritchard2019} Web site URL, that is, https://www.basissetexchange.org.

In the current implementation, after the program reads a given input
file, the required one- and two-electron atomic integrals are calculated
using the original numerical quadrature developed in the HONDO program \citep{Dupuis1989}. 
Recall that ERIs over molecular orbitals are required, but only two-index $\mathcal{J}_{pq}$ and
$\mathcal{K}_{pq}$ integrals are necessary due to our approximation
for the 2RDM. We expand the NOs in a fixed atomic
basis set, namely,
\begin{equation}
\varphi_{p}\left(\mathbf{r}\right)=\sum_{\upsilon=1}^{\mathrm{N}_{B}}\mathcal{C}_{\upsilon p}\zeta_{\upsilon}\left(\mathbf{\mathbf{r}}\right)\,,\quad p=1,\ldots,\mathrm{N}_{B}\label{lcao}
\end{equation}
where $\mathrm{N}_{B}$ is the number of basis functions. Accordingly,
Coulomb and exchange integrals are calculated as
\begin{equation}
\mathcal{J}_{pq}=\sum_{\mu,\upsilon=1}^{\mathrm{N}_{B}}\Gamma_{\mu\upsilon}^{p}\mathcal{J}_{\mu\upsilon}^{q}\,,\quad\mathcal{K}_{pq}=\sum_{\mu,\upsilon=1}^{\mathrm{N}_{B}}\Gamma_{\mu\upsilon}^{p}\mathcal{K}_{\mu\upsilon}^{q}\label{JKpq}
\end{equation}
where
\begin{equation}
\Gamma_{\mu\upsilon}^{p}=\mathcal{C}_{\mu p}\mathcal{C}_{\upsilon p},
\end{equation}

\begin{equation}
\mathcal{J}_{\mu\upsilon}^{q}=\sum_{\eta,\delta=1}^{\mathrm{N}_{B}}\Gamma_{\eta\delta}^{q}\left\langle \mu\eta|\upsilon\delta\right\rangle ,\,\mathcal{K}_{\mu\upsilon}^{q}=\sum_{\eta,\delta=1}^{\mathrm{N}_{B}}\Gamma_{\eta\delta}^{q}\left\langle \mu\eta|\delta\upsilon\right\rangle \label{Gpmn}
\end{equation}

From Eqs. (\ref{JKpq})-(\ref{Gpmn}), we can see that the four-index
transformation of the ERIs scales as $\mathrm{N}_{B}^{5}$. In the
occupancy optimization, this operation is carried out once for fixed
orbitals, however, in the orbital optimization it is necessary to
perform this transformation every time the orbitals change to generate
the matrix $\mathcal{F}$, which is a time-consuming process. The
parallel implementation of this part of the code substantially improved
the performance of our program \citep{Matito2013}. It should be noted
that for some approximations where the two-electron cumulant can be
factorized is possible to reduce this formal scaling to $\mathrm{N}_{B}^{4}$
by summing over orbital indexes separately \citep{Giesbertz2016,MITXELENA2019}.

\begin{figure*}
\caption{\label{FlowChart} Flowchart in DoNOF}

\noindent \centering{}\includegraphics[scale=0.37]{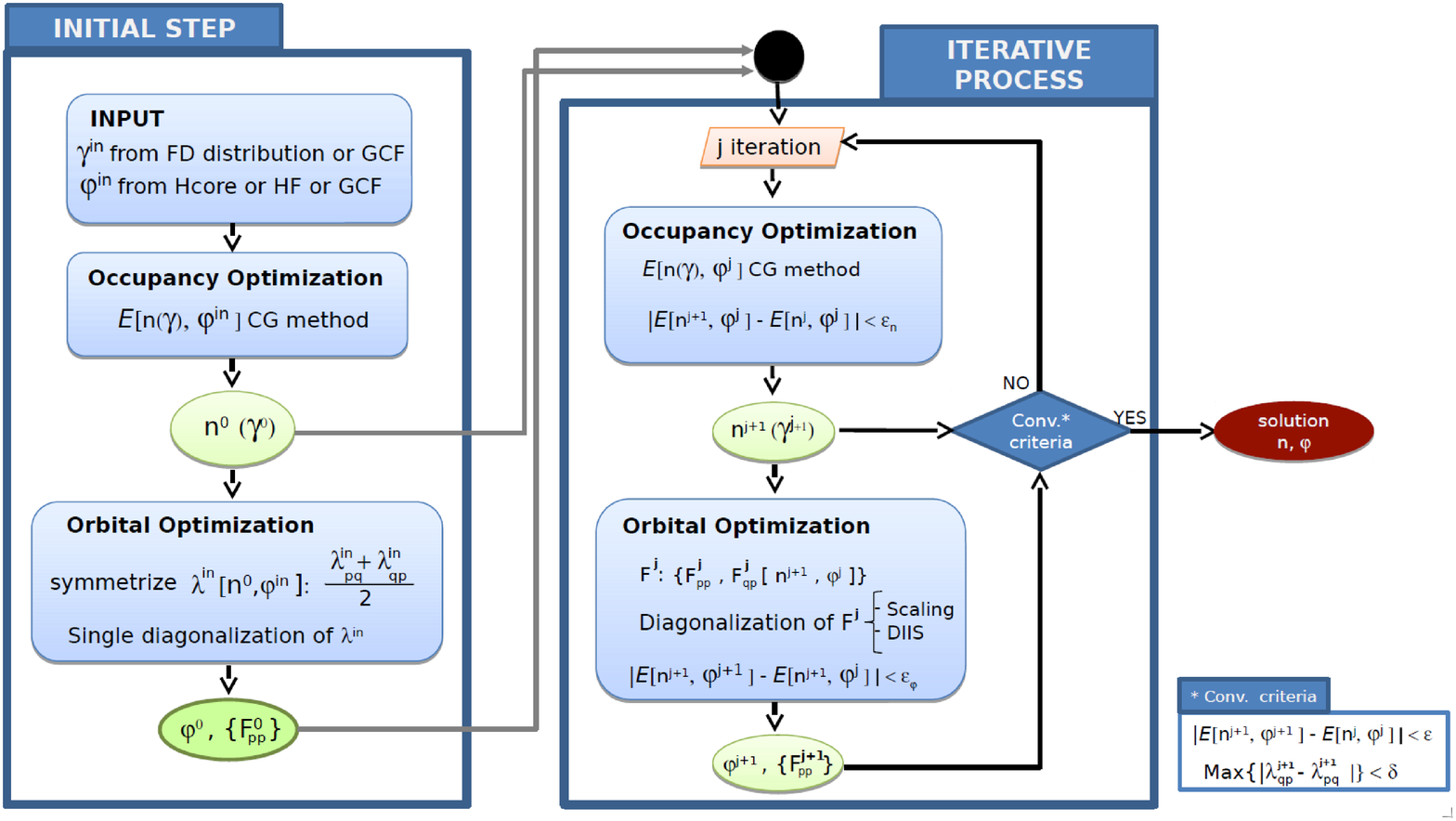}
\end{figure*}

Fig. \ref{FlowChart} depicts the flowchart in DoNOF. The procedure
is divided into two fundamental steps. An initial one, which main
objective is to generate the diagonal elements of the symmetric matrix
$\mathcal{F}$, and the iterative step where it is optimized by occupancies
and orbitals separately.

In the initial step, the guess for molecular orbitals $\left\{ \varphi_{p}^{in}\right\} $
is generated directly from the core Hamiltonian or after a HF calculation.
An energy optimization with respect to \textcolor{black}{the auxiliary
$\gamma$-variables is performed using the CG or L-BFGS methods.}
For this purpose, initial $\left\{ \gamma_{p}^{in}\right\} $ corresponding
to a Fermi-Dirac (FD) distribution of the ONs is generally used. For
the ONs $\left\{ n_{p}^{0}\right\} $ obtained according to Eqs. (\ref{soo})
and (\ref{woo}) from resulting $\left\{ \gamma_{p}^{0}\right\} $,
a single diagonalization of the symmetrized $\lambda$-matrix with
$\left\{ \varphi_{p}^{in}\right\} $, $\left(\lambda_{pq}^{in}+\lambda_{qp}^{in}\right)/2$,
is carried out that affords $\left\{ \varphi_{p}^{0}\right\} $ and
a suitable start guess $\left\{ \mathcal{F}_{pp}^{0}\right\} $.

In the iterative step, $\gamma^{j+1}$ is obtained for fixed NOs $\varphi^{j}$
by the unconstrained optimization of $E\left[n\left(\gamma\right),\varphi^{j}\right]$.
Then, the symmetric matrix $\mathcal{F}^{j}$ is formed with the resulting
diagonal elements of the previous step, and the off-diagonal elements
given by Eq. (\ref{Fockian}) using $n^{j+1}\left(\gamma^{j+1}\right)$
and $\varphi^{j}$. After diagonalization of $\mathcal{F}^{j}$, new
NOs $\varphi^{j+1}$ and diagonal elements $\left\{ \mathcal{F}_{pp}^{j+1}\right\} $
are obtained.

Three criteria of energy convergence are used, one for occupancy optimization,
another for orbital optimization and finally a global one. Each criterion
consists of monitoring the electronic energy of each iteration and
requires that two successive values \LyXZeroWidthSpace \LyXZeroWidthSpace differ
by no more than a threshold, namely,

\begin{equation}
\left|E\left[n^{j+1},\varphi^{j}\right]-E\left[n^{j},\varphi^{j}\right]\right|<\varepsilon_{n}
\end{equation}
\begin{equation}
\left|E\left[n^{j+1},\varphi^{j+1}\right]-E\left[n^{j+1},\varphi^{j}\right]\right|<\varepsilon_{\varphi}
\end{equation}
\begin{equation}
\left|E\left[n^{j+1},\varphi^{j+1}\right]-E\left[n^{j},\varphi^{j}\right]\right|<\varepsilon
\end{equation}

It is worth noting that we do not determine whether the new NOs $\varphi^{j+1}$
(ONs $n^{j+1}$) are the same as previous $\varphi^{j}$ ($n^{j}$).
A robust criterion that prevails over the three energy criteria is
to assess the symmetry of the matrix of the Lagrange multipliers,
namely, the maximum of the off-diagonal differences to be less than
$\delta$:
\begin{equation}
Max\left\{ \left|\lambda_{qp}^{j+1}-\lambda_{pq}^{j+1}\right|\right\} <\delta
\end{equation}

To help convergence, the scaling of $\mathcal{F}$ is used as well
as the DIIS technique explained above. If the convergence criteria
are reached, then the solution represented by optimal $\left\{ n_{p},\varphi_{p}\right\} $
can be employed to calculate other quantities of interest.

\section{\label{sec: MolPro}Molecular Properties}

At the present stage, the program computes molecular properties such
as vertical ionization potentials by means of the extended Koopmans\textquoteright{}
theorem \citep{Piris2012}, Mulliken population analysis \citep{Mulliken1955},
\textcolor{black}{and electric dipole, quadrupole and octupole moments}
\citep{Mitxelena2016}. We provide the RDMs in the atomic and molecular
bases if necessary, and a wavefunction (WFN) file for performing further
quantitative and visual analyses of molecular systems.

The 1RDM and Lagrangian cannot be simultaneously brought to the diagonal
form in approximate NOFs. DoNOF works on the NO representation, where
the 1RDM is diagonal, but $\lambda$ is \LyXZeroWidthSpace \LyXZeroWidthSpace only
a symmetric matrix. The program can provide a complementary canonical
representation \citep{Piris2013} of the one-electron picture in molecules
where $\lambda$ is diagonal, but not the 1RDM. This equivalent representation
affords delocalized molecular orbitals adapted to the symmetry of
the molecule. Combining both representations we have the whole picture
concerning the ONs and orbital energies in NOF theory. Currently both
types of orbitals can be displayed using the DoNOF output file which
can be read by the GMOLDEN programm \citep{MOLDEN} duly modified
by us.

In the case of PNOF5, an APSG with the expansion coefficients explicitly expressed
by the ONs can generate \citep{Piris2013e} the functional too. Our
code offers the option of knowing the expansion coefficients of the
APSG generating wavefunction. In addition, the treatment of an extended
system at a fractional cost of the entire calculation can also be
achieved using a thermodynamic fragment energy method in the context
of NOF theory \citep{Lopez2015a}.

Geometry optimization is, together with single-point energy calculations,
the most used procedure in electronic structure calculations. DoNOF
computes the analytic energy gradients \citep{Mitxelena2017} with
respect to nuclear motion without resorting to linear response theory.
It is worth noting that our implementation of the geometry optimization
can also be applied to non-singlet systems as long as the spin-multiplet
formalism mentioned above is used\textcolor{black}{{} \citep{Mitxelena2020b}.
}This efficient computation of analytical gradients, analogous to
gradient calculations at the HF level of theory, allows to locate
and characterize the critical points on the energy surface.

In contrast to first-order energy derivatives, the calculation of
the second-order analytic derivatives requires the knowledge of NOs
and ONs at the perturbed geometry \citep{Mitxelena2018}. Thus, since
a set of coupled-perturbed equations must be solved to analytically
obtain the Hessian in NOF theory, the corresponding computational
cost increases dramatically. Consequently, in the current implementation,
the Hessian is obtained by numerical differentiation of analytical
gradients \citep{MITXELENA2019}.

\section{\label{sec: DynCorr}Dynamic Correlation}

Current NOFs based on electron pairing take into account most of the
non-dynamical effects, and also an important part of the dynamical
electron correlation corresponding to the intrapair interactions.
Consequently, electron-pairing-based NOFs produce \citep{Piris2017,Piris2013e}
results that are in good agreement with accurate wavefunction-based
methods for small systems, where electron correlation effects are
almost entirely intrapair. When the number of pairs increases, NOF
values deteriorate especially in those regions where dynamic correlation
prevails. It is therefore mandatory to add the inter-space dynamic
electron correlation to improve calculations.

The second-order Møller--Plesset (MP2) perturbation theory is the
simplest way of properly incorporating dynamic correlation effects.
In DoNOF, the recently proposed \citep{Piris2019,Piris2017,Piris2018b}
\textcolor{black}{NOF-MP2 method can be used to obtain a good balance
}of both static (non-dynamic) and dynamic electron correlation. The
zeroth-order Hamiltonian is constructed from a closed-shell-like Fock
operator that contains a HF density matrix with doubly ($2n_{g}=2$)
and singly ($2n_{g}=1$) occupied orbitals. The electronic energy
is then calculated by the expression
\begin{equation}
E=\tilde{E}_{hf}+E^{corr}=\tilde{E}_{hf}+E^{sta}+E^{dyn}\label{Etotal}
\end{equation}

\noindent where $\tilde{E}_{hf}$ is the HF-like energy obtained with
the NOs,
\begin{equation}
\tilde{E}_{hf}=2\sum\limits _{g=1}^{\mathrm{N}_{\Omega}}\mathcal{H}_{gg}+\sum_{f,g=1}^{\mathrm{N}_{\Omega}}\left(2\mathcal{J}_{fg}-\mathcal{K}_{fg}\right)-\sum_{g=\frac{\mathrm{N_{II}}}{2}+1}^{\mathrm{N}_{\Omega}}\frac{\mathcal{J}_{gg}}{4}\label{HFL}
\end{equation}

\noindent $E^{sta}$ is the sum of the static intra-space and inter-space
correlation energies,
\begin{equation}
\begin{array}{c}
E^{sta}=\sum\limits _{g=1}^{\mathrm{N_{II}/2}}\sum\limits _{q\neq p}\sqrt{\Lambda_{q}\Lambda_{p}}\,\Pi_{qp}\mathcal{\,K}_{pq}\qquad\qquad\\
\\
-4\sum\limits _{f\neq g}^{\mathrm{\mathrm{N}_{\Omega}}}\sum\limits _{p\in\Omega_{f}}\sum\limits _{q\in\Omega_{g}}\Phi_{q}^{2}\Phi_{p}^{2}\mathcal{K}_{pq}
\end{array}\label{Esta}
\end{equation}

\noindent and $E^{dyn}$ is obtained from a modified second-order
MP2 correction,
\begin{equation}
E^{dyn}=\sum\limits _{g,f=1}^{\mathrm{\mathrm{N}_{\Omega}}}\;\sum\limits _{p,q>\mathrm{N}_{\Omega}}^{\mathrm{N}_{B}}A_{g}A_{f}\left\langle gf\right|\left.pq\right\rangle \left[2T_{pq}^{gf}\right.\left.-T_{pq}^{fg}\right]\label{E2}
\end{equation}
where
\begin{equation}
A_{g}=\left\{ \begin{array}{c}
1\,,\quad1\leq g\leq\frac{\mathrm{N_{II}}}{2}\qquad\\
\frac{\mathrm{1}}{2},\:\frac{\mathrm{N_{II}}}{2}<g\leq\mathrm{N}_{\Omega}
\end{array}\right.
\end{equation}

\noindent $\mathrm{N}_{B}$ is the number of basis functions, and
$T_{pq}^{fg}$ are the amplitudes of the doubly excited configurations
obtained by solving modified equations for the MP2 residuals \citep{Piris2018b}.
An important feature of the method is that double counting is avoided
by taking the amount of static and dynamic correlation in each orbital
as a function of its occupancy. In Eq. (\ref{Esta}), for instance,
$\Lambda_{p}=1-\left|1-2n_{p}\right|$ goes from zero for empty or
fully occupied orbitals to one if the orbital is half occupied.

It is convenient to take into account the inter-pair static correction
in the reference-used NOF from the outset, thus preventing the ONs
and NOs from suffering an inter-pair non-dynamic influence, however
small, in the dynamic correlation domains. This led us to correlate
the motion of electrons with parallel and opposite spins belonging
to different subspaces ($\Omega_{f}\neq\Omega{}_{g}$) as
\begin{equation}
E_{fg}=\sum\limits _{p\in\Omega_{f}}\sum\limits _{q\in\Omega_{g}}\left[n_{q}n_{p}\left(2\mathcal{J}_{pq}-\mathcal{K}_{pq}\right)-4\Phi_{q}^{2}\Phi_{p}^{2}\mathcal{K}_{pq}\right]\label{Efg_s}
\end{equation}
The functional (\ref{EPNOF7}), (\ref{Eg}), and (\ref{Efg_s}) instead
of Eq. (\ref{Efg}) was called static PNOF7 (PNOF7s) for multiplets
\citep{Piris2018b}. Note that the difference between
the PNOF7 and PNOF7s functionals lies in the last term of $E_{fg}$
that correlates electrons with opposite spins. In the case of PNOF7s,
the quadratic dependence of the product $4\Phi_{q}^{2}\Phi_{p}^{2}$
practically eliminates this term when the ONs are close to zero or
one, that is, when dynamic correlation is present. On the other hand,
its value is maximum when the orbitals are half occupied, that is,
static correlation dominates, coinciding with the PNOF7 values of
the $E_{fg}$ in Eq. (\ref{Efg}).

\section{\label{sec: Examples}Illustrative example: water molecule}

The water molecule is probably the most studied system in quantum
chemistry. We have accurate experimental data at our disposal for
$H_{2}O$, therefore it is an ideal candidate to test our code and
display the capabilities of DoNOF. In this section, we employ the
correlation-consistent basis set series cc-pVXZ (X = D, T, Q) developed
by Dunning and coworkers \citep{Dunning1989} obtained from use of
the basis set exchange software \citep{Pritchard2019}. For comparison,
we report the experimental data as well as other accurate theoretical
calculations taken from the NIST CCCBDB database \citep{NIST}.

Convergence was accomplished in all cases for the three energy criteria
$\varepsilon_{n},\,\varepsilon_{\varphi}=10^{-10},\,\varepsilon=10^{-8}$,
and a tolerance $\delta=10^{-4}$, in accordance to the symmetry criterion
of the matrix of the Lagrange multipliers. A curious fact is that,
in the case of a lower energy criterion $\varepsilon$, the accuracy
achieved in most cases exceeds the required value when the $\delta$-tolerance
is reached, confirming that it is the strongest of all the criteria
discussed in the subsection \ref{subsec:Procedure}.

\begin{table}
\noindent \centering{}\caption{\label{Table_ONs}Occupation numbers of the water molecule at the
experimental equlibrium geometry ($R_{OH}=0.9578\textrm{Å}$, $\varangle HOH=104.5\text{º}$)
and atomic dissociation. The cc-pVDZ basis set was employed. $\mathrm{N}_{g}=4$.\medskip{}
}
\begin{tabular}{c|ccc|ccc}
\multirow{2}{*}{MO} & \multicolumn{3}{c|}{Experimental Geometry} & \multicolumn{3}{c}{Atomic Dissociation}\tabularnewline
 & PNOF5 & PNOF7s & PNOF7 & PNOF5 & PNOF7s & PNOF7\tabularnewline
\hline 
1 & 2.000 & 2.000 & 2.000 & 2.000 & 2.000 & 2.000\tabularnewline
2 & 1.980 & 1.980 & 1.971 & 1.000 & 1.000 & 1.000\tabularnewline
3 & 1.980 & 1.980 & 1.971 & 1.000 & 1.000 & 1.000\tabularnewline
4 & 1.992 & 1.992 & 1.988 & 1.993 & 1.991 & 1.976\tabularnewline
5 & 1.992 & 1.992 & 1.988 & 1.993 & 1.991 & 1.976\tabularnewline
6 & 0.000 & 0.000 & 0.000 & 0.000 & 0.001 & 0.002\tabularnewline
7 & 0.001 & 0.001 & 0.001 & 0.001 & 0.001 & 0.003\tabularnewline
8 & 0.007 & 0.007 & 0.011 & 0.005 & 0.007 & 0.016\tabularnewline
9 & 0.000 & 0.000 & 0.002 & 0.001 & 0.001 & 0.003\tabularnewline
10 & 0.000 & 0.000 & 0.000 & 0.000 & 0.001 & 0.002\tabularnewline
11 & 0.000 & 0.000 & 0.002 & 0.001 & 0.001 & 0.003\tabularnewline
12 & 0.001 & 0.001 & 0.001 & 0.001 & 0.001 & 0.003\tabularnewline
13 & 0.007 & 0.007 & 0.011 & 0.005 & 0.007 & 0.016\tabularnewline
14 & 0.001 & 0.001 & 0.001 & 0.000 & 0.000 & 0.000\tabularnewline
15 & 0.002 & 0.002 & 0.003 & 0.000 & 0.000 & 0.001\tabularnewline
16 & 0.017 & 0.017 & 0.025 & 1.000 & 1.000 & 0.998\tabularnewline
17 & 0.001 & 0.001 & 0.001 & 0.000 & 0.000 & 0.001\tabularnewline
18 & 0.001 & 0.001 & 0.001 & 0.000 & 0.000 & 0.001\tabularnewline
19 & 0.017 & 0.017 & 0.025 & 1.000 & 1.000 & 0.998\tabularnewline
20 & 0.002 & 0.002 & 0.003 & 0.000 & 0.000 & 0.001\tabularnewline
21 & 0.000 & 0.000 & 0.000 & 0.000 & 0.000 & 0.000\tabularnewline
22 & 0.000 & 0.000 & 0.000 & 0.000 & 0.000 & 0.000\tabularnewline
23 & 0.000 & 0.000 & 0.000 & 0.000 & 0.000 & 0.000\tabularnewline
24 & 0.000 & 0.000 & 0.000 & 0.000 & 0.000 & 0.000\tabularnewline
25 & 0.000 & 0.000 & 0.000 & 0.000 & 0.000 & 0.000\tabularnewline
\end{tabular} 
\end{table}

We begin by selecting the minimum value of $\mathrm{N}_{g}$ required
to correctly describe the electron pairs that make up the system,
which in the case of water are five pairs. Our electron-pair-based
functionals are not capable of recovering the entire dynamic correlation,
so we must resort to perturbative corrections if we want to obtain
significant total energies. Consequently, it is convenient to reduce
the value of $\mathrm{N}_{g}$ considering only the ONs that do not
exceed a certain threshold, e.g. 0.01. Small ONs \LyXZeroWidthSpace \LyXZeroWidthSpace are
known to contribute solely to dynamic correlation and slow down convergence
by making the surface where minimization is performed flatter.

We must recall that we assumed an equal number $\mathrm{N}_{g}$ for
all subspaces that is determined by the basis set used. In general,
the best strategy is to consider the smallest basis set to determine
which are the orbitals that determine the static correlation in different
significant geometries, and thus define the minimum required value
of $\mathrm{N}_{g}$.

Table \ref{Table_ONs} shows the ONs obtained with PNOF5, PNOF7s and
PNOF7 at the experimental equilibrium geometry. In principle, a geometry
close to equilibrium, e.g., the geometry optimized at the HF theory
level, is adequate. The table also includes the ONs \LyXZeroWidthSpace \LyXZeroWidthSpace in
the atomic dissociation. For the latter, the distance between the
oxygen atom and each hydrogen atom was taken equal to 1000 $\textrm{Å}$.
The $\mathrm{N}_{g}$ value corresponding to the cc-pVDZ basis set
was found to be four.

Inspection of the data collected in this table reveals that for $H_{2}O$
the typical bonding-anti-bonding orbital scheme is adequate. Indeed,
the ONs of the weakly occupied NOs 8, 13, 16 and 19, determine that
the value of $\mathrm{N}_{g}=1$ is sufficient. Furthermore, taking
into account how full the first orbital is, we can also freeze its
occupancy at 2. Note that the NOs together with their occupancies
adapt to the problem during the optimization process. We have not
rearranged the ONs since our objective is to highlight the coupling
scheme adopted to form the subspaces (see Fig. \ref{fig1}).

In Table \ref{Table_ONs-1}, we can find the ONs obtained by establishing
$\mathrm{N}_{g}=1$. The results confirm that no significant differences
are obtained, although they are better adapted to the symmetry of
the system when the number of optimized $\gamma$-variables is reduced.
Hereinafter, all our results refer to this bonding-anti-bonding orbital
scheme.

\begin{table}
\noindent \centering{}\caption{\label{Table_ONs-1}Occupation numbers of the water molecule at the
experimental equlibrium geometry ($R_{OH}=0.9578\textrm{Å}$, $\varangle HOH=104.5\text{º}$)
and atomic dissociation. The cc-pVDZ basis set was employed. $\mathrm{N}_{g}=1$.\medskip{}
}
\begin{tabular}{c|ccc|ccc}
\multirow{2}{*}{MO} & \multicolumn{3}{c|}{Experimental Geometry} & \multicolumn{3}{c}{Atomic Dissociation}\tabularnewline
 & PNOF5 & PNOF7s & PNOF7 & PNOF5 & PNOF7s & PNOF7\tabularnewline
\hline 
1 & 2.00000 & 2.00000 & 2.00000 & 2.00000 & 2.00000 & 2.00000\tabularnewline
2 & 1.98183 & 1.98158 & 1.97575 & 1.00000 & 1.00000 & 1.00000\tabularnewline
3 & 1.98183 & 1.98158 & 1.97575 & 1.00000 & 1.00000 & 1.00000\tabularnewline
4 & 1.99306 & 1.99297 & 1.99051 & 1.99464 & 1.99242 & 1.98303\tabularnewline
5 & 1.99306 & 1.99297 & 1.99051 & 1.99464 & 1.99242 & 1.98303\tabularnewline
6 & 0.00694 & 0.00703 & 0.00949 & 0.00536 & 0.00758 & 0.01697\tabularnewline
7 & 0.00694 & 0.00703 & 0.00949 & 0.00536 & 0.00758 & 0.01697\tabularnewline
8 & 0.01817 & 0.01842 & 0.02425 & 1.00000 & 1.00000 & 1.00000\tabularnewline
9 & 0.01817 & 0.01842 & 0.02425 & 1.00000 & 1.00000 & 1.00000\tabularnewline
\end{tabular}
\end{table}

\begin{figure*}
\noindent \centering{}\includegraphics[scale=0.45]{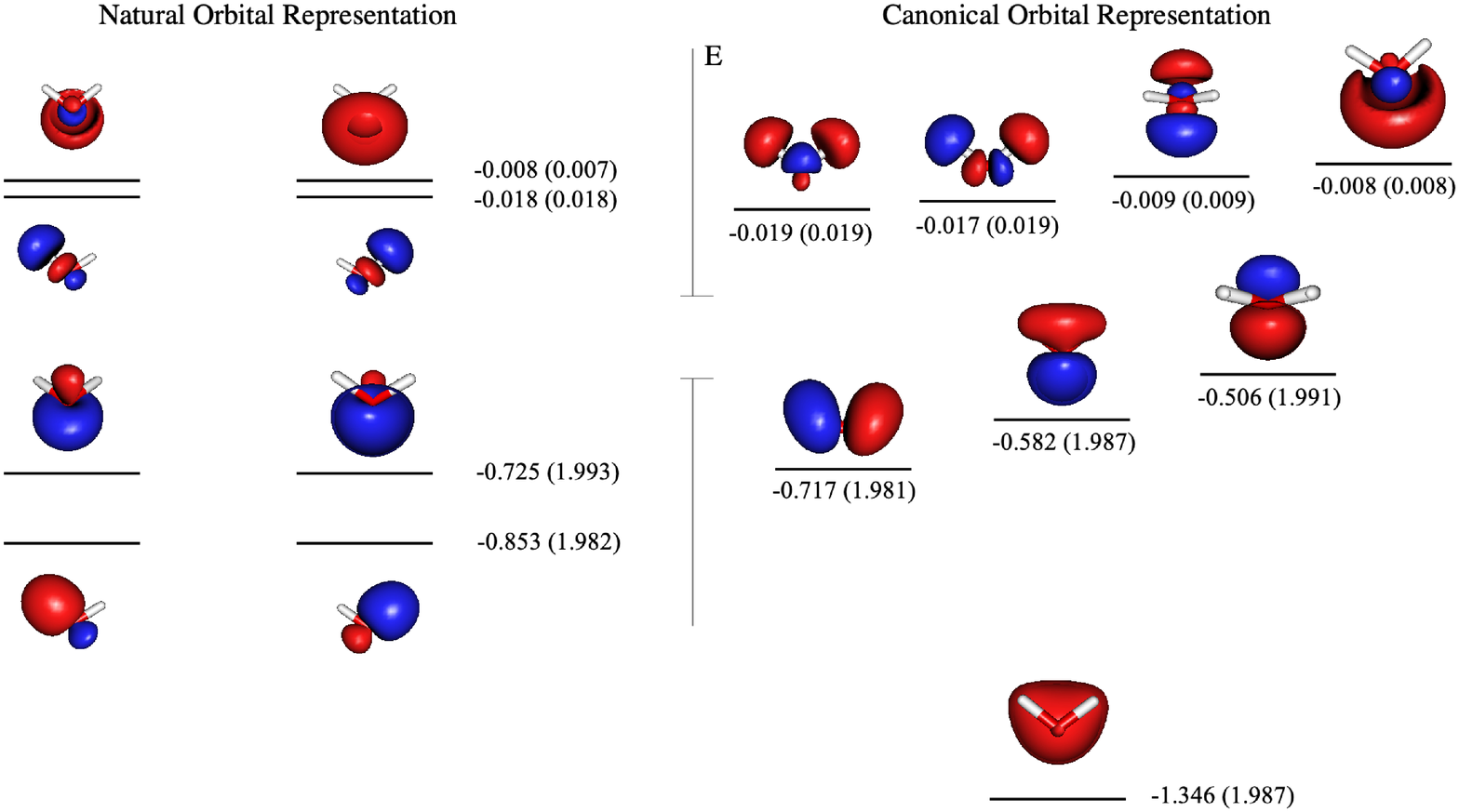}\caption{\label{Fig-Orbitals} Valence molecular orbitals calculated using
PNOF7s/cc-pVDZ level of theory at the equilibrium geometry. Corresponding
diagonal Lagrange multipliers in Hartrees, and 1RDM diagonal elements
in parentheses, are also reported}
\end{figure*}

The PNOF7s valence NOs are shown in Fig. \ref{Fig-Orbitals} at the
experimental equilibrium geometry. Recall that in the NO representation,
the ONs are well defined but not the orbital energies. DoNOF provides
the complementary canonical orbital (CO) representation included on
the right side of the figure, where orbital energies correspond to
the diagonal elements of the matrix of Langrange multipliers. Regrettably,
we cannot assign ONs to the orbitals in the CO representation. In
fact, the maximum 1RDM off-diagonal element is 0.028 at the PNOF7s/cc-pVDZ
level of theory. When 1RDM off-diagional elements are small enough,
the negative values of CO energies below the Fermi level provides
a good approximation to the ionization potentials \citep{Piris2013}. 

In Fig. \ref{Fig-Orbitals}, we observe that NOs closely match the
image emerging from the chemical bonding arguments: the O atom has
$sp3$ hybridization, two of orbitals are used to bind to H atoms,
leading to two degenerate OH $\sigma$ bonds, and the remaining two
are degenerated lone pairs. On the other hand, in the CO representation,
the orbitals obtained adapt to the molecular symmetry and resemble
those obtained by the usual molecular orbital theories, for example,
HF. Both representations provide the complete picture of occupancies
and orbital energies.

The next step is to optimize the geometry of the water molecule for
each level of theory. We use HF geometries as starting points to PNOF
optimizations. Table \ref{Table_OPtGeo} shows the errors in OH bond
lengths and HOH bond angle with respect to the experimental structural
data. According to the errors, PNOF5 and PNOF7s provide ground-state
equilibrium bond-distances comparable to those of the MP2 and CCSD(T),
whereas PNOF7 deviates slightly from these accurate wavefunction-based
methods. Note that the results obtained with all three functionals
for the HOH bond angle are excellent.

\begin{table}[h]
\noindent \centering{}\caption{\label{Table_OPtGeo}Differences of the optimized geometry with respect
to experimental data ($R_{OH}=0.9578\textrm{Å}$, $\varangle HOH=104.5\text{º}$)\medskip{}
}
\begin{tabular}{c|cc|cc|cc}
\multirow{2}{*}{Method} & \multicolumn{2}{c|}{cc-pVDZ} & \multicolumn{2}{c|}{cc-pVTZ} & \multicolumn{2}{c}{cc-pVQZ}\tabularnewline
 & $R_{OH}$($\textrm{Å}$) & $\varangle\left(\lyxmathsym{\textdegree}\right)$ & $R_{OH}$($\textrm{Å}$) & $\varangle\left(\lyxmathsym{\textdegree}\right)$ & $R_{OH}$($\textrm{Å}$) & $\varangle\left(\lyxmathsym{\textdegree}\right)$\tabularnewline
\hline 
PNOF5 & 0.0072 & -1.2 & 0.0005 & $\,$0.3 & -0.0007 & $\,$0.6\tabularnewline
PNOF7s & 0.0075 & -1.2 & 0.0007 & $\,$0.3 & -0.0004 & $\,$0.6\tabularnewline
PNOF7 & 0.0134 & -1.7 & 0.0066 & -0.2 & $\,$0.0052 & $\,$0.1\tabularnewline
MP2 & 0.0071 & -2.6 & 0.0012 & -1.0 & -0.0001 & -0.5\tabularnewline
CCSD(T) & 0.0086 & -2.5 & 0.0017 & -0.9 & $\,$0.0001 & -0.4\tabularnewline
\end{tabular}
\end{table}

A comparison between harmonic vibrational frequencies obtained by
using PNOF5, PNOF7s, PNOF7, MP2, and CCSD(T), with respect to experimental
fundamentals is displayed in Table \ref{Table_Freq}. According to
reported values, all functionals show good agreement with MP2 and
CCSD(T). The largest error of 180 cm$^{-1}$ corresponds to the third
vibrational mode computed with PNOF7 which, as mentioned above, can
take an excess of static correlation in the equilibrium region that
corresponds to the dynamic inter-pair correlation. The frequencies
\LyXZeroWidthSpace \LyXZeroWidthSpace obtained with MP2 indicate that
the NOF-MP2 method could lead to excellent results, a fact that we
have verified in dimers \citep{Piris2017}.

\begin{table}[h]
\noindent \centering{}\caption{\label{Table_Freq}Differences in harmonic vibrational frequencies
(cm$^{-1}$) with respect to experimental data (3832, 1649, 3943)\medskip{}
}
\begin{tabular}{c|ccc|ccc|ccc}
\multirow{1}{*}{Method} & \multicolumn{3}{c|}{cc-pVDZ} & \multicolumn{3}{c|}{cc-pVTZ} & \multicolumn{3}{c}{cc-pVQZ}\tabularnewline
\hline 
PNOF5 & -37 & 104 & -65 & -8 & 86 & -34 & 11 & 79 & -16\tabularnewline
PNOF7s & -45 & 105 & -66 & -9 & 130 & 19 & 34 & 128 & 30\tabularnewline
PNOF7 & -162 & 82 & -180 & -125 & 56 & -146 & -128 & 5 & -142\tabularnewline
MP2 & 20 & 29 & 28 & 24 & 3 & 34 & 23 & -6 & 35\tabularnewline
CCSD(T) & -12 & 41 & -17 & 8 & 19 & 2 & 13 & 10 & 9\tabularnewline
\end{tabular}
\end{table}

\begin{figure*}
\begin{centering}
\includegraphics[scale=0.47]{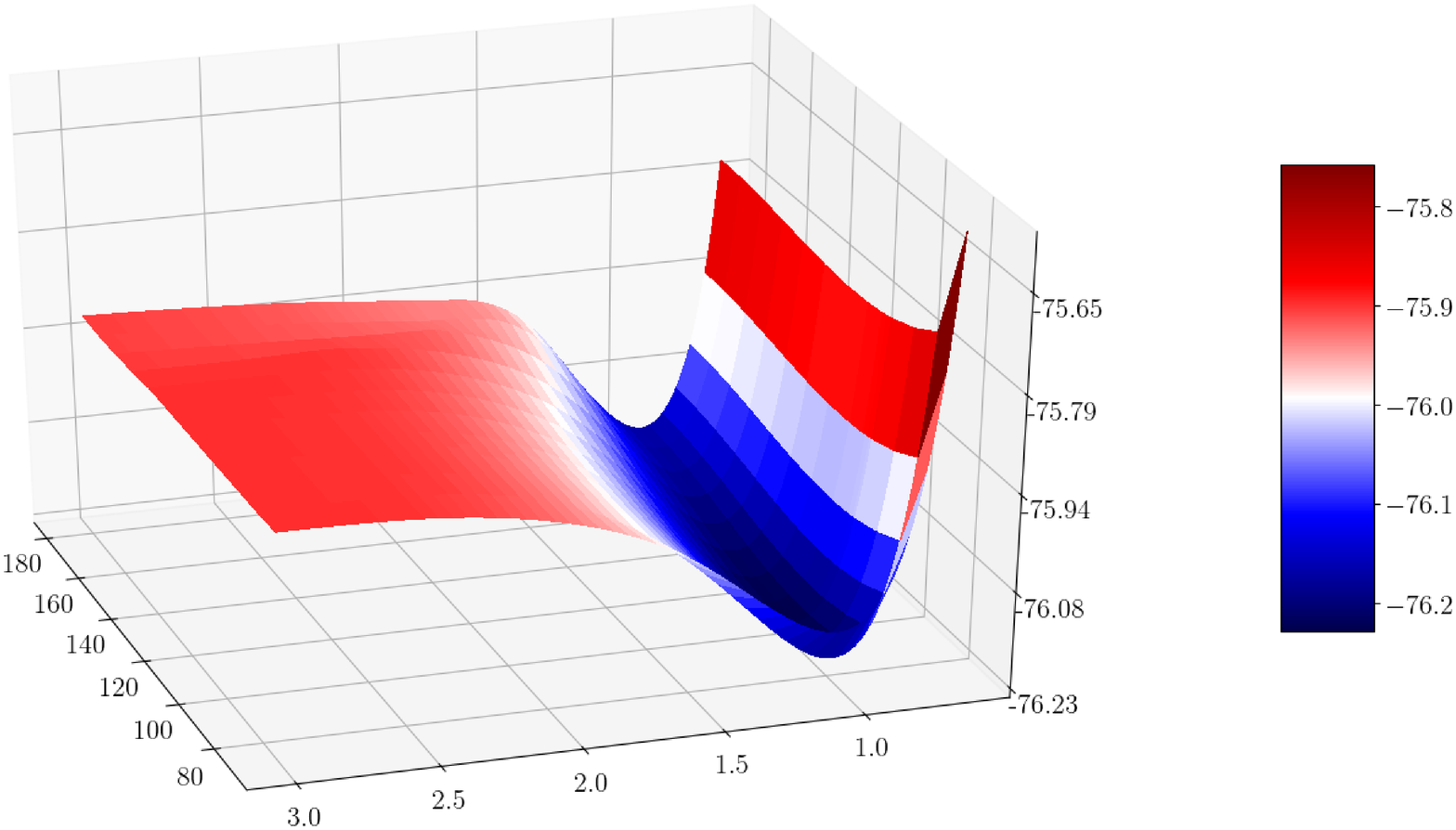}
\par\end{centering}
\caption{\label{fig:SymDisH2O}Symmetric dissociation of the water molecule
at the NOF-MP2/cc-pVDZ level of theory. The minimum is at $R_{OH}=0.966\textrm{ Å}$,
$\varangle HOH=102.7\lyxmathsym{\protect\textdegree}$}
\end{figure*}

Unfortunately, analytical gradients for the NOF-MP2 method are not
available in DoNOF. It is a purpose for the near future, but unlike
PNOF gradients, NOF-MP2 gradients require the linear response to nuclear
motion. Nevertheless, we have scanned the energy with respect to the
OH distance and the HOH angle to obtain the potential energy surface
(PES). This procedure is computationally expensive, so we only show
the PES at the NOF-MP2/cc-pVDZ level of theory (Fig. \ref{fig:SymDisH2O}).

The minimum ($R_{OH}=0.966\,\textrm{Å}$, $\varangle HOH=102.7\lyxmathsym{\textdegree}$)
confirms that NOF-MP2 tends to lengthen bond distances and decrease
bond angles with respect to PNOF7s. In the case of water, these trends
would improve the PNOF7s marks with respect to the experimental values
\LyXZeroWidthSpace \LyXZeroWidthSpace when the basis set increases.

It is worth noting that the NOFs used here as well as the NOF-MP2
method are size-consistent. This property implies that the energy
of a water molecule obtained when the oxygen and hydrogens are infinitely
separated is identical to the sum of the atomic energies. Furthermore,
our NOFs are known to be the only ones that provide an integer number
of electrons in the dissociated atoms when calculating Mulliken populations
\citep{Matxain2011}. Both properties have been verified numerically
in this study for $R_{OH}=1000\,\textrm{Å}$.

In Table \ref{Table_BindingE}, we have collected the energies ($D_{e}$)
required to disassemble a water molecule into its constituent atoms
calculated at different levels of theory. Recall that MP2 and CCSD(T)
fail at the dissociation limit of water, so these $D_{e}$ values
\LyXZeroWidthSpace \LyXZeroWidthSpace have been calculated as the
difference between the energy of the three atoms and the energy of
water at equilibrium. In addition, MP2 and CCSD(T) employ unrestricted
formulations to characterize spin-uncompensated systems.

\begin{table}[H]
\noindent \centering{}\caption{\label{Table_BindingE}Theoretical binding energies ($D_{e}$) in
kcal/mol at 0K}
\medskip{}
\begin{tabular}{c|ccc}
\multirow{1}{*}{Method} & \multicolumn{1}{c}{cc-pVDZ} & \multicolumn{1}{c}{cc-pVTZ} & \multicolumn{1}{c}{cc-pVQZ}\tabularnewline
\hline 
MP2 & 210.8 & 228.5 & 234.0\tabularnewline
CCSD(T) & 208.8 & 225.0 & 229.9\tabularnewline
NOF-MP2 & 222.2 & 242.2 & 247.4\tabularnewline
\end{tabular}
\end{table}

The values of $D_{e}$ \LyXZeroWidthSpace \LyXZeroWidthSpace obtained
with our NOFs are not reported in Table \ref{Table_BindingE} since
they underestimate $D_{e}$ by approximately 50 kcal/mol due to the
aforementioned lack of an important part of the dynamic correlation
in the equilibrium region. Conversely, NOF-MP2 adequately recovers
the dynamic correlation and yields more reasonable values \LyXZeroWidthSpace \LyXZeroWidthSpace of
the binding energy. We can see that these values, however, are higher
than those obtained with the MP2 and CCSD(T) methods.

In the case of singlets at the equilibrium geometry, NOF-MP2 provides
total energies similar to those of the MP2 method \citep{Piris2018b},
whereas CCSD(T) can further reduce energy with the triple contributions.
Accordingly, the differences in $D_{e}$ may be associated with uneven
descriptions of the dissociated oxygen atom in its triplet state.
Indeed, the oxygen total spin is conserved with NOF-MP2 in contrast
to the MP2 and CCSD(T) methods. This symmetry violation leads to lower
energy values \LyXZeroWidthSpace \LyXZeroWidthSpace for the triplet
state. Table \ref{Table_OxygenGap} shows the differences of the singlet-triplet
gap with respect to the experiment for the oxygen atom calculated
with the three methods. From the Table, we can conclude that the best
description of the singlet-triplet gap is given by the NOF-MP2 method,
whereas MP2 significantly overestimates the energy of the triplet
state. The CCSD(T) method compensates for the spin violation in the
triplet state with the best description of the electron correlation
in the singlet state.
\begin{center}
\begin{table}[H]
\noindent \centering{}\caption{\label{Table_OxygenGap}Energy differences (kcal/mol) in the singlet-triplet
gap of the oxygen atom with respect to the experiment (45.4 kcal/mol)}
\medskip{}
\begin{tabular}{c|ccc}
\multirow{1}{*}{Method} & \multicolumn{1}{c}{cc-pVDZ} & \multicolumn{1}{c}{cc-pVTZ} & \multicolumn{1}{c}{cc-pVQZ}\tabularnewline
\hline 
MP2 & 23.4 & 21.1 & 19.8\tabularnewline
CCSD(T) & 6.9 & 5.7 & 5.0\tabularnewline
NOF-MP2 & 2.8 & 0.1 & -0.8\tabularnewline
\end{tabular}
\end{table}
\par\end{center}

Next, let us discuss two more properties that we can obtain with our
code in the corresponding optimized geometries, namely, ionization
potentials and electrostatic moments.

Table \ref{Table_IPs} lists the differences of the lowest vertical
ionization potential (LVIP) with respect to the experimental values
obtained by PNOF5, PNOF7s, PNOF7 using the extended Koopmans' theorem
(EKT) \citep{Piris2012}. For comparison, the Koopmans' theorem (KT)
LVIP has also been included. It has long been recognized that, in
general, there is an excellent agreement between the KT LVIPs and
the experimental data because of the fortuitous cancellation of the
electron correlation and orbital relaxation effects. However, in the
case of water, the KT greatly overestimates the LVIP.
\begin{center}
\begin{table}[H]
\noindent \begin{centering}
\caption{\label{Table_IPs}Differences in eV of the lowest vertical ionization
potentials with respect to the experimental data (12.600 eV). PNOF
values were obtained by means of the extended Koopmans' theorem (EKT)\medskip{}
}
\begin{tabular}{c|ccc}
\multirow{1}{*}{Method} & \multicolumn{1}{c}{cc-pVDZ} & \multicolumn{1}{c}{cc-pVTZ} & \multicolumn{1}{c}{cc-pVQZ}\tabularnewline
\hline 
Koopmans' & 0.848 & 1.173 & 1.265\tabularnewline
PNOF5 & 0.461 & 0.801 & 0.896\tabularnewline
PNOF7s & 0.458 & 0.798 & 0.892\tabularnewline
PNOF7 & 0.394 & 0.726 & 0.815\tabularnewline
$\Delta$NOFMP2{*} & \multicolumn{1}{c}{-0.225} & \multicolumn{1}{c}{0.399} & \multicolumn{1}{c}{0.584}\tabularnewline
\end{tabular}\medskip{}
\par\end{centering}
\centering{}{*} The PNOF7s optimized geometry was used
\end{table}
\par\end{center}

A survey of Table \ref{Table_IPs} reveals that PNOF-EKT values improve
those obtained by KT, a predominant trend previously observed in many
systems \citep{Piris2012}. The EKT considers the electron correlation
accounted by the specific functional, but neglects the orbital relaxation
in the (N-1)-state. In order to include the dynamic correlation and
take into account the orbital relaxation, we have calculated the LVIP
as the energy difference between the positive ion in its doublet state
and the neutral water molecule, at the NOF-MP2 level of theory using
PNOF7s optimized geometries. These values \LyXZeroWidthSpace \LyXZeroWidthSpace appear
in the table under the $\Delta$NOFMP2 notation. We can perceive an
improvement in the $\Delta$NOFMP2 values approaching the experimental
LVIP. However, neither method produces better agreement with the experiment
when the size of the basis set improves.

Due to its increased electronegativity, the oxygen atom in the water
molecule attracts the electrons shared with the hydrogen atoms in
the covalent bonds. As a result, the O atom acquires a partial negative
charge, while the H atoms have a partial positive charge. Additionally,
the electron lone-pair on the oxygen forces the molecule to assume
a bent structure. The latter avoids the cancellation of polar OH bonds,
causing the entire water molecule to become polarized.

Tables \ref{Table_DipMom} and \ref{Table_QuadMom} show the differences
with respect to the experiment in the calculated electric dipole and
quadrupole moments at the HF, PNOF5, PNOF7s, PNOF7 and CCSD levels
of theory, respectively. The latter were computed with respect to
the center of mass using the corresponding optimized geometry for
each theoretical method\textcolor{black}{. Since the traceless definition
of the quadrupole moment is employed, two components are sufficient
to determined it. The octupole moment has been excluded from this
study, since the latter is only significant when the lower order electric
moments are zero, e.g. in tetrahedral molecules \citep{Mitxelena2016}.}
\begin{center}
\textcolor{black}{}
\begin{table}[h]
\noindent \centering{}\textcolor{black}{\caption{\label{Table_DipMom}Differences in Debye of the water dipole moment
calculated at the optimized geometry with respect to the experimental
data (1.855 Debye)\medskip{}
}
}%
\begin{tabular}{c|ccc}
Method & \textcolor{black}{cc-pVDZ} & \textcolor{black}{cc-pVTZ} & \textcolor{black}{cc-pVQZ}\tabularnewline
\hline 
\textcolor{black}{HF} & \textcolor{black}{0.189} & \textcolor{black}{0.133} & \textcolor{black}{0.110}\tabularnewline
\textcolor{black}{PNOF5} & \textcolor{black}{0.174} & \textcolor{black}{0.128} & \textcolor{black}{0.102}\tabularnewline
\textcolor{black}{PNOF7s} & \textcolor{black}{0.173} & \textcolor{black}{0.128} & \textcolor{black}{0.101}\tabularnewline
\textcolor{black}{PNOF7} & \textcolor{black}{0.161} & \textcolor{black}{0.117} & \textcolor{black}{0.091}\tabularnewline
\textcolor{black}{CCSD} & \textcolor{black}{0.128} & \textcolor{black}{0.072} & \textcolor{black}{0.055}\tabularnewline
\end{tabular}
\end{table}
\par\end{center}

\textcolor{black}{Inspection of the data collected in these tables
reveals that PNOFs produce molecular electric moments comparable to
those of CCSD. Interestingly, in the case of the water molecule, the
HF results are in line with those obtained by methods that include
electron correlation, although this is generally not the case \citep{Mitxelena2016}.}
According to Tables \textcolor{black}{\ref{Table_DipMom} and \ref{Table_QuadMom}},
the charge distribution of molecules is adequately described using
any of the PNOF approaches. However, in view of the aforementioned
missing interpair dynamical correlation, it is important to study
the electrical properties using the NOF-MP2 method. A work in this
direction is underway.

\textcolor{black}{}
\begin{table}[H]
\noindent \centering{}\textcolor{black}{\caption{\label{Table_QuadMom}Differences in Buckingham of the water quadrupole
moment components calculated at the optimized geometry with respect
to the experimental data ($Q_{xx}=-2.500,Q_{yy}=2.630$)\medskip{}
}
}%
\begin{tabular}{c|cc|cc|cc}
\multirow{2}{*}{Method} & \multicolumn{2}{c|}{\textcolor{black}{cc-pVDZ}} & \multicolumn{2}{c|}{\textcolor{black}{cc-pVTZ}} & \multicolumn{2}{c}{\textcolor{black}{cc-pVQZ}}\tabularnewline
 & \textcolor{black}{$Q_{xx}$} & \textcolor{black}{$Q_{yy}$} & \textcolor{black}{$Q_{xx}$} & \textcolor{black}{$Q_{yy}$} & \textcolor{black}{$Q_{xx}$} & \textcolor{black}{$Q_{yy}$}\tabularnewline
\hline 
\textcolor{black}{HF} & \textcolor{black}{0.504} & \textcolor{black}{-0.324} & \textcolor{black}{0.351} & \textcolor{black}{-0.087} & \textcolor{black}{0.295} & \textcolor{black}{-0.011}\tabularnewline
\textcolor{black}{PNOF5} & \textcolor{black}{0.559} & \textcolor{black}{-0.363} & \textcolor{black}{0.358} & \textcolor{black}{-0.103} & \textcolor{black}{0.291} & \textcolor{black}{-0.016}\tabularnewline
\textcolor{black}{PNOF7s} & \textcolor{black}{0.559} & \textcolor{black}{-0.364} & \textcolor{black}{0.358} & \textcolor{black}{-0.104} & \textcolor{black}{0.291} & \textcolor{black}{-0.017}\tabularnewline
\textcolor{black}{PNOF7} & \textcolor{black}{0.493} & \textcolor{black}{-0.311} & \textcolor{black}{0.365} & \textcolor{black}{-0.118} & \textcolor{black}{0.294} & \textcolor{black}{-0.025}\tabularnewline
\textcolor{black}{CCSD} & \textcolor{black}{0.560} & \textcolor{black}{-0.502} & \textcolor{black}{0.368} & \textcolor{black}{-0.205} & \textcolor{black}{0.285} & \textcolor{black}{-0.086}\tabularnewline
\end{tabular}
\end{table}

\section{\label{sec: summary}Summary}

In this work, we have presented the DoNOF program for computational
chemistry calculations based on the natural orbital functional (NOF)
theory, where the electronic structure is described in terms of the
natural orbitals (NOs) and their occupation numbers (ONs).

In section \ref{sec:Introduction}, we discussed how NOF approximations
can be useful to quantum chemistry as an alternative formalism to
both density functional and wavefunction methods. Special emphasis
was placed on the need for functional N-representability and the conservation
of the total spin of the system under study.

In section \ref{sec:NOF}, we reviewed the fundamental aspects of
the electron-pairing-based NOF theory for multiplets. The reconstructions
developed in our group for the two-particle reduced density matrix
in terms of the ONs were briefly presented. They led to the appearance
of different versions of PNOF that can provide a correct description
of systems with a multiconfigurational nature, one of the biggest
challenges for DFs.

Our self-consistent algorithm for optimizing an energy functional
with respect to the NOs and their ONs was introduced in section \ref{sec: Energy-minimization}.
We described how the constrained nonlinear programming problem for
ONs can be treated as unconstrained optimization, with corresponding
computational savings. The specific techniques that help obtain convergence
in the orbital optimization were discussed, namely the direct inversion
of the iterative subspace (DIIS) extrapolation technique, and the
variable scaling of the symmetric matrix subject to iterative diagonalizations. 
It is important to note that, although these techniques
have confirmed their practical value, the number of iterations is
still very high, so more convergence techniques must be implemented
in the code. Depending on the desired precision, computational times
can be long, limiting the calculations to small molecules with extended
basis sets or modestly sized molecules with modest basis sets. 
An overview of the computational procedure in DoNOF is given at the
end of this section.

Section \ref{sec: MolPro} was dedicated to molecular properties that
can be calculated with DoNOF once the NOF solution is reached. The
orbital-invariant formulation of the NOF-MP2 method was briefly presented
in section \ref{sec: DynCorr}. The latter allows us to remedy the
lack of the interpair dynamic correlation inherent in $\mathcal{JKL}$-functionals
developed by us. The single-reference NOF-MP2 method improves the
absolute energies over the PNOF values \LyXZeroWidthSpace \LyXZeroWidthSpace by
bringing them closer to the values \LyXZeroWidthSpace \LyXZeroWidthSpace obtained
by accurate wavefunction-based methods.

The capabilities of DoNOF were discussed in section \ref{sec: Examples}
through the water molecule. This included analysis of molecular orbitals,
ionization potential, electric moments, optimized geometry, harmonic
vibrational frequencies, binding energy, and potential energy surface
of the $H_{2}O$ symmetric dissociation. Consequently,
DoNOF can calculate the energy and a variety of properties of any
isolated molecule in its ground state, regardless of the total spin
value of the system. It is especially important when one encounters
near degeneracies, such as those that occur during bond-breaking processes,
near transition states in chemical reactions or transition metal compounds.
However, at present calculations of excited electronic states, the
influence of a solvent, or relativistic effects cannot be performed.
It is anticipated that these features will be implemented into DoNOF
in the near future.

To extend the calculations with DoNOF to larger systems,
in addition to making the iterative process more effective and efficient,
new parallel implementations of other computationally demanding tasks
are needed. On the other hand, the local spatial nature of NOs could
be exploited to achieve a better scaling of the code with respect
to the number of basis functions. It has been shown with other methods
that an important source of computational savings is the neglect of
correlation for pairs of very distant localized orbitals. Consequently,
an improved scaling could be achieved by adopting a preselection procedure
for the density matrices involved that will reduce the number of two-electron
integrals.

We have shown the solution to longstanding problems in NOF theory,
however new algorithms and capabilities are welcome to be included
in the code. New parallel implementations of the DoNOF program tasks
are also needed. We believe that DoNOF will benefit the community
of quantum physicists and chemists, mainly those who start in the
development of new functionals, and these in turn will improve the
code by bringing new functionalities.

\medskip{}

\noindent \textbf{\textcolor{black}{Acknowledgments:}}\textcolor{black}{{}
The authors greatly appreciate }Dr. Eduard Matito's\textcolor{black}{{}
parallel} implementation of the four-index transformation of the electron
repulsion integrals that substantially improved the performance of
DoNOF. \textcolor{black}{The authors also appreciate for technical
and human support provided by IZO-SGI SGIker of UPV/EHU and European
funding (ERDF and ESF). Financial support comes from MCIU/AEI/FEDER,
UE (PGC2018-097529-B-100) and }Eusko Jaurlaritza (Ref. IT1254-19)\textcolor{black}{. }

\end{document}